\RequirePackage{fix-cm}
\documentclass[natbib]{svjour3}                     
\smartqed  
\usepackage{graphicx}
\usepackage{mathptmx}      
\usepackage{amssymb}
\usepackage{amsmath}
\usepackage{hyperref}
\journalname{Experimental astronomy}
\begin{document}

\title{Detecting weak beryllium lines with CUBES\thanks{The authors acknowledge support by the National Science Centre, Poland, project 2018/31/B/ST9/01469}
}

\titlerunning{Detecting weak Be lines with CUBES}        

\author{Rodolfo~Smiljanic         \and
        Andr\'e~R.~da Silva       \and 
        Riano~E.~Giribaldi
}

\authorrunning{Smiljanic, da Silva \& Giribaldi} 

\institute{R. Smiljanic, A.~R.~da Silva, R.~E.~Giribaldi  \at
           Nicolaus Copernicus Astronomical Center, Polish Academy of Sciences, ul. Bartycka 18, 00-716, Warsaw, Poland\\
              \email{rsmiljanic@camk.edu.pl}           
}

\date{Received: date / Accepted: date}
\maketitle

\begin{abstract}
Beryllium is a light element with one single stable isotope, $^9$Be, which is a pure product of cosmic-ray spallation in the interstellar medium. Beryllium abundances in late-type stars can be used in studies about evolutionary mixing, Galactic chemical evolution, planet engulfment, and the formation of globular clusters. Some of these uses of Be abundances figure among the science cases of the Cassegrain U-Band Efficient Spectrograph (CUBES), a new near-UV low- and medium-resolution spectrograph under development for the Very Large Telescope. Here, we report on a study about beryllium abundances in extremely metal-poor stars in the context of the phase A of CUBES. Our motivation is to understand the limits for the detection of weak lines in extremely metal-poor stars of low Be abundances. We analyze simulated CUBES observations, performed in medium-resolution mode, based on synthetic spectra for four mock stars with [Fe/H] $\leq$ $-$3.0. We find that detecting the Be lines is possible in certain cases, but is very challenging and requires high signal-to-noise ratio. Depending on the atmospheric parameters of the target stars, and if signal-to-noise per pixel of about 400 can be achieved, it should be possible to detect Be abundances between $\log$(Be/H) $-$13.1 and $-$13.6, with a typical uncertainty of $\pm$ 0.15 dex. Using CUBES, the required data for such studies can be obtained for stars that are fainter by two magnitudes with respect to what is possible with current instrumentation.  

\end{abstract}

\section{Introduction}\label{intro}

The near ultraviolet (near-UV) spectral region (which we define here as the region between $\sim$ 3000--4000 \AA) contains the only absorption lines of beryllium that can be studied in the spectra of late-type stars, when using ground-based instrumentation. These are the lines located at 3130.423 and 3131.067 \AA\, which originate from the 2s $^2$S$_{1/2}$ $\rightarrow$ 2p $^2$P$_{1/2,3/2}$ transitions in singly-ionized beryllium \citep{2005PhyS...72..309K}.

Beryllium has atomic number Z = 4 and a single stable isotope, $^9$Be. Interestingly, the Be atoms that we observe in stellar photospheres are a pure product of cosmic-ray spallation, being formed in the interstellar medium from the break up of CNO nuclei \citep{2018ARNPS..68..377T}. As became clear from early works such as \citet{1988A&A...193..193R} and \citet{1992Natur.357..379G}, in old, metal-poor stars the relation between the abundance of Be and the stellar metallicity\footnote{When referring to metallicity we mean the iron abundance, [Fe/H], which is the usual rough indicator of the total metal content of a star. Abundances in the bracket notation for two elements A and B, i.e. [A/B], mean the difference between the logarithm of the ratio of the abundances by number in a star and the logarithm of the same ratio in the Sun: [A/B] = $\log$[N(A)/N(B)]$_{\star}$ - $\log$[N(A)/N(B)]$_{\odot}$. The abundances by number are given in a scale where the number of hydrogen atoms is N(H) = 10$^{12}$.} is linear. Such a linear relation implies that Be is produced as a primary element, i.e.\ its production rate is independent of metallicity \citep{2012A&A...542A..67P}. Primary elements are those produced directly from H and He. The only way to obtain Be that behaves as a primary element is to assume that the Galactic cosmic rays that are important for the Be nucleosynthesis are made of accelerated CNO nuclei which then collide with material of the interstellar medium \citep{1992ApJ...401..584D,1997ApJ...488..338D}. Such a fundamental observation about the evolution of Be in the Galaxy helps in the understanding of the sources and sites from where Galactic cosmic-rays are accelerated \citep[see e.g.][and references therein]{2019ApJS..245...30L}.

The study of Be abundances in old, metal-poor stars ([Fe/H] $<\,-$1.00) started with the few objects analyzed by \citet{1984ESASP.218..197M}, \citet{1984A&A...139..394M}, \citet{1988A&A...193..193R}, \citet{1991ApJ...378...17G}, and \citet{1992ApJ...388..184R}. Near-UV sensitive spectrographs at 8-10m class-telescopes allowed the determination of Be abundances in larger samples, of the order of 150 metal-poor stars down to [Fe/H] $\sim\,-$3.00 \citep{2009A&A...499..103S,2009MNRAS.392..205T,2011ApJ...743..140B}. Several science cases can benefit if the sample of metal-poor stars analyzed for Be abundances increases and is extended to include stars of even lower metallicities. 

In this context, the Cassegrain U-Band Efficient Spectrograph (CUBES, see Zanutta et al., CUBES Phase-A design overview, ExA, in press as part of the CUBES Special Issue) is expected to make an important contribution. CUBES is planned to be installed at the Very Large Telescope (VLT) of the European Southern Observatory (ESO), in Cerro Paranal, Chile. The CUBES instrument was first discussed in \citet{2014SPIE.9147E..09B} and \citet{2014Ap&SS.354..191B}. Future giant telescopes, like the Extremely Large Telescope (ELT) of ESO, will be optimized for observations in the red/infrared. Therefore, it was realized that a UV optimized spectrograph at the VLT would remain the best option for studies requiring wavelength coverage bluer than 4000 \AA \citep{2014Ap&SS.354..121P,2016SPIE.9908E..9JE}. The current CUBES design envisions a high-efficiency spectrograph able to operate in low- (R $\sim$ 6000) and medium-resolution (R $\sim$ 24\,000) dedicated to the near-UV spectral region (Zanutta et al.). The two resolution modes are obtained with the use of two different image slicers (Calcines et al., Design of the VLT-CUBES image slicers, ExA, in press as part of the CUBES Special Issue) in combination with first-order transmission gratings (Zeitner et al., ExA, in press as part of the CUBES Special Issue). The science case for such an instrument has been summarized before in \citet{2018SPIE10702E..2EE} and \citet{2020SPIE11447E..60E}, and is further discussed in several contributions to this Special Issue.

In this work, we present our simulations of the expected performance of CUBES regarding the detection of weak Be lines in faint, extremely metal-poor stars. We build upon, and expand, previous simulations reported in \cite{2014Ap&SS.354...55S} by incorporating the recent updates in the CUBES design. This work is divided as follows. In Section \ref{sec:science}, we summarize the science case that motivates the observation of the Be lines in the spectra of metal-poor stars. In Section \ref{sec:linelist}, we discuss the experimental data and codes used to generate the simulated stellar spectra. Section \ref{sec:limits} presents the analysis that attempts to determine the Be abundances of the simulated observations and a discussion about the limits for the detection of Be lines using CUBES spectra. Finally, in Section \ref{sec:conclusions} we summarize our findings and highlight a few important caveats that one should keep in mind when interpreting our simulations. 

\section{Beryllium in metal-poor stars}\label{sec:science}

The models of inhomogeneous (or stochastic) chemical enrichment of the Galactic halo built by \citet{1999ApJ...522L.125S} and \citet{2001ApJ...549..303S} suggested that Be abundances in the early Galaxy should have a smaller scatter than the abundances of elements produced by stellar nucleosynthesis (such as oxygen or iron). This led to the suggestion that Be abundances could be used as a good cosmic clock for the early Galaxy \citep{2000IAUS..198..425B,2001ApJ...549..303S}. To test this, \citet{2004A&A...426..651P,2007A&A...464..601P} investigated Be abundances in turn-off stars of two globular clusters. The abundances were converted into ages using chemical evolution models computed by \citet{2002ApJ...566..252V}. The comparison of these values with the cluster ages derived from isochrone fitting seemed to support the idea of Be as a cosmochronometer, even though the uncertainty of the abundances was high. We expect that CUBES will allow an increase in the number of globular clusters that can be used for such analysis (see Giribaldi et al., ExA, in press as part of the CUBES Special Issue).

The possibility to use Be abundances as a time scale is particularly attractive for studies of early Galaxy formation and evolution. Recently, \textit{Gaia} data \citep{2016A&A...595A...1G,2018A&A...616A..10G} revealed strong evidence that the Milky Way suffered a major merger between 8 to 11 Gyr ago \citep{2018MNRAS.478..611B,2018Natur.563...85H,2018ApJ...863..113H}. Hints of this event had already been identified in previous data, e.g. \citet{2000AJ....119.2843C}, but it was the \textit{Gaia} data that revealed the full picture of the system involved in the merger. It is now becoming apparent that the evolution of Be abundances as a function of metallicity is different in the distinct systems that built up the Galactic halo  \citep{2020MNRAS.496.2902M,2021A&A...646A..70S}. The first hints that Be could be used to separate accreted and in-situ stars, in a diagram of [O/Fe] as a function of log(Be/H), were found by \citet{2005A&A...436L..57P} and later confirmed by \citet{2009A&A...499..103S} and \citet{2011ApJ...738L..33T}. Independently on whether or not Be is indeed a good cosmic clock, it does offers a different ``nucleosynthetic axis'' to study the early chemical evolution of the Milky Way.

There has also being a push towards attempting to determine Be abundances in several extremely metal-poor stars with [Fe/H] $\lesssim~-$3.00 \citep{2000A&A...362..666P,2000A&A...364L..42P,2006ApJ...641.1122B,2009ApJ...701.1519R,2013ApJ...773...33I,2014ApJ...790...34P,2019A&A...624A..44S,2021A&A...646A..70S,2022MNRAS.510.5362E}. One motivation for this was to investigate the possibility of primordial production of substantial amounts of $^9$Be. Standard primordial (or Big Bang) nucleosynthesis predicts the formation of tiny amounts of Be, at the level of $\log$(Be/H) $\sim$ $-$17.6 \citep{2014JCAP...10..050C}. Several alternative scenarios could result in an enhanced primordial Be abundance of up to $\log$(Be/H) $\sim$ $-$15 \citep{1989ApJ...336L..55B,1997ApJ...488..515O,2008JCAP...11..020P,2014ApJS..214....5K,2017IJMPE..2641004K}. Nevertheless, the Be abundances currently detected in extremely metal-poor stars are at the level of $\log$(Be/H) $\sim$ $-$13.5 with a few upper limits down to $\log$(Be/H) $\sim$ $-$14.2 \citep{2014ApJ...790...34P,2019A&A...624A..44S,2021A&A...646A..70S}. This is still one to two orders of magnitude above what could be produced by non-standard primordial nucleosynthesis, and thus not at the level that could help to constrain such models.

It is thus important to expand investigations of Be abundances in metal-poor stars, both for tracing the early star forming conditions of the Galaxy and for constraining any possible deviation from standard Big Bang nucleosynthesis. The extremely metal-poor stars (with [Fe/H] $<$ $-$3.0) useful for these analyses are mostly located at far distances, at the Galactic halo, and are thus faint. Moreover, the Be lines for these stars are very small and therefore spectra of high signal-to-noise ratio (SNR) are needed for a proper analysis. Essentially, all such stars that could be easily analysed with current instrumentation have already been observed. For example, \citet{2021A&A...646A..70S} report on new near-UV spectra for star BPS BS 16968-061, with [Fe/H] = $-$2.98 and $V$ = 13.22 mag, obtained with the Ultraviolet and Visual Echelle Spectrograph \citep[UVES,][]{2000SPIE.4008..534D} at the VLT. A total of 20 hours of observation resulted in SNR $\sim$ 200 per pixel in the region around the Be lines. Only a new generation of highly efficient near-UV spectrographs, like CUBES, can help to observe increased samples of such stars using shorter observing times.

\section{Simulated spectra}\label{sec:linelist}

To simulate the CUBES observations of metal-poor stars, we computed synthetic spectra covering not only the near-UV but also wavelengths up to about 7000 \AA. This is needed because we rely on the redder wavelengths to compute the stellar magnitudes in the $V$ band. The work to compile and test the atomic and molecular line list needed for these calculations is described in the companion paper by Giribaldi et al. (in this Special Issue). Here, we describe only our literature investigation which aimed to compile the best available data for the two Be spectral lines, including for the calculation of their hyperfine structure.

\subsection{Hyperfine structure of the Be lines}

\begin{table}
	\caption{Adopted data for the Be resonance lines.}
	\label{tab:1}       
	\begin{tabular}{lll}
		\hline\noalign{\smallskip}
		wavelength (\AA) & $\log(gf)$ & $\sigma$.$\alpha$  \\
		\noalign{\smallskip}\hline\noalign{\smallskip}
		3130.423 & $-$0.178 & 123.212 \\
		3131.067 & $-$0.479 & 123.212 \\
		\noalign{\smallskip}\hline
	\end{tabular}
\end{table}

The simulations discussed here, have been performed with an updated line list with respect to what has been used in our recent work \citep{2021A&A...646A..70S}. The recent literature was scanned for modern results of the atomic data needed to compute the Be spectral lines. In particular, our line list now includes the hyperfine splitting of the Be lines. We thus take the opportunity to describe these updates.

We adopt absolute transition frequencies measured in \cite{2015PhRvL.115c3002N} and \citet{2017ApPhB.123...15K}, which have uncertainties smaller by almost two orders of magnitude in comparison with the previous values \citep{1985PhRvA..31.2711B}. We note that the experimental fine structure splitting is in very good agreement with the modern theoretical value \citep{2015PhRvA..92a2513P}, the difference being negligible for our purposes. The transition wavelengths, in \AA, are listed in Table \ref{tab:1}.

For the two Be lines, the critical compilation of transition probabilities by \citet{2010JPCRD..39a3101F} recommends results of the accurate theoretical calculations by  \citet{1998PhRvA..57.1652Y}. We investigated more recent results of lifetimes and oscillator strengths \citep{2009PhRvA..80d2511T,2013PhRvA..87c2502S,2020PhRvA.102b2817S} but found no change that could affect the determination of stellar chemical abundances. The adopted values of $\log(gf)$ are given in Table \ref{tab:1}.

For the broadening due to collisions with hydrogen atoms, calculations based on the theory developed by \cite{1991MNRAS.253..549A} and \citet{1998MNRAS.300..863B} are available. The line broadening cross sections are tabulated with values computed at a velocity of 10000 m s$^{-1}$, and the cross-sections are assumed to vary according to a power law on the velocity of the form $v^{-\alpha}$. The cross section ($\sigma$) and the $\alpha$ parameter computed by \cite{2000MNRAS.311..535B} are listed in Table \ref{tab:1}.

For the hyperfine (HFS) structure, we adopt results from the theoretical calculations of \cite{2008PhRvA..78a2513Y}. The magnetic dipole (A) and electric quadrupole (B) HFS constants needed to compute the energy shifts of each sublevel are listed in Table \ref{tab:2}. Other calculations reported in the literature give slightly different values for some of the constants \citep{2009PhRvA..79c2510P,2014PhRvA..89c2510P}. We have tested these values and found that the differences are at a level that causes no effect on the HFS needed for synthesizing stellar spectra.

The hyperfine line splitting is a result of the interaction between the electronic angular momentum, J, and the nuclear angular momentum, I, resulting in a total angular momentum F. The nuclear angular momentum for $^9$Be is I = 3/2. The selection rules allow transitions between levels with $\Delta$F = $\pm$1,0 but not between two levels with F = 0. This results in the splitting of the 3130.423 \AA\ line into six components and of the 3131.067 \AA\ line into four components. The energy shifts and line strengths were computed with equations that can be found in texts like \citet{woodgate1970elementary} and \citet{2006sham.book..253E}, and with help of the code provided by \citet{2011arXiv1102.5125M}. The wavelengths and $\log(gf)$ of the hyperfine transitions are given in Table \ref{tab:3}.

\begin{table}
	\caption{Constants needed to compute the energy shifts of the hyperfine sublevels \citep{2008PhRvA..78a2513Y}.}
	\label{tab:2}       
	\begin{tabular}{lrrr}
		\hline\noalign{\smallskip}
		  & $^2$S$_\frac{1}{2}$ & $^2$P$_{\frac{1}{2}}$  & $^2$P$_{\frac{3}{2}}$\\
		\noalign{\smallskip}\hline\noalign{\smallskip}
		A (MHz) & $-$625.11 & $-$117.925 & $-$1.016 \\
		B (MHz) & -- & -- &   $-$2.281 \\
		\noalign{\smallskip}\hline
	\end{tabular}
\end{table}

\begin{table}
	\caption{Hyperfine structure of the Be lines.}
	\label{tab:3}       
	\begin{tabular}{lll}
		\hline\noalign{\smallskip}
		wavelength (\AA) & $\log(gf)$ & F$_l$ -- F$_u$  \\
		\noalign{\smallskip}\hline\noalign{\smallskip}
3130.425 & $-$1.382  &  1 -- 0 \\
3130.425 & $-$0.984  &  1 -- 1 \\ 
3130.425 & $-$0.984  &  1 -- 2 \\
3130.421 & $-$1.683  &  2 -- 1 \\
3130.421 & $-$0.984  &  2 -- 2 \\
3130.421 & $-$0.537  &  2 -- 3 \\
3131.069 & $-$1.683  &  1 -- 1 \\
3131.070 & $-$0.984  &  1 -- 2 \\
3131.065 & $-$0.984  &  2 -- 1 \\
3131.066 & $-$0.984  &  2 -- 2 \\
\noalign{\smallskip}\hline
\end{tabular}
\end{table}

\subsection{Simulated observations}

Synthetic stellar spectra were computed with the code {\sf Turbospectrum} \citep{2012ascl.soft05004P} using the MARCS one-dimensional model atmospheres \citep{2008A&A...486..951G} under the local thermodynamical equilibrium (LTE) approximation. The spectra were computed between 3000-7000 \AA\ on steps of 0.02 \AA.

In total, spectra of four stars were simulated. We chose to simulate spectra of two main-sequence stars with ($T_\mathrm{eff}$, $\log~g$) = (6300 K, 4.30) and metallicities [Fe/H] = $-$3.0 and $-$3.5. These are parameters similar to the stars analysed in \citet{2021A&A...646A..70S}. We also simulated spectra for two subgiants with ($T_\mathrm{eff}$, $\log~g$) = (5600 K, 3.40) and the same two metallicity values, [Fe/H] = $-$3.0 and $-$3.5. These temperature and $\log~g$ values are the same of the star analyzed in \citet{2019A&A...624A..44S}. All stars were simulated with microturbulence velocity ($\xi$) of 1.50 km s$^{-1}$ and using $\alpha$-enhanced models with [$\alpha$/Fe] = +0.4. With these choices we approximately bracket the parameter space for the extremely metal-poor stars that one would select to observe. 

In addition, for each of the four stars, 20 spectra with different abundances of Be were computed where one of these 20 was always simulated without the Be lines. For the stars with [Fe/H] = $-$3.0, the other 19 spectra have log(Be/H) that varies between $-$13.8 and $-$12.9 in steps of 0.05 dex (i.e.\ [Be/Fe] varying from $-$0.12 to +0.78). For the stars with [Fe/H] = $-$3.5, the 19 spectra have log(Be/H) that varies between $-$14.1 and $-$13.2 in steps of 0.05 dex (i.e.\ [Be/Fe] varying from +0.08 to +0.98). Thus, our initial set of simulations includes 80 spectra.

To understand the properties of the observed spectra, we make use of the CUBES end-to-end simulator (E2E) and the CUBES exposure time calculator (ETC) described in Genoni et al. (The CUBES Instrument model and simulation tools, ExA, in press as part of the CUBES Special Issue). The E2E simulates the effects of the atmosphere, the fore-optics, the image slicers, the spectrograph, and the detector on a real astronomical observation. According to the E2E, the current CUBES design provides R $\sim$ 23\,000 with a sampling of $\sim$2.35 px (pixels of 0.058 \AA) at 3131 \AA, the region of the Be lines.

To estimate the SNR of the observations, we used the ETC assuming observations for a bright ($V$ = 12.5 mag, in the Vega system) and a faint ($V$ = 14 mag) case. The SNR was estimated for observations with airmass = 1.2 and with a seeing of 0.9 arcsec at 5500 \AA\ (which corresponds to about 1.0 arcsec at 3100 \AA). The exposure time was set to 3000s. This time was chosen to simulate an observing block of one hour, as is typical for service mode observations in Paranal, assuming 10 min of overhead, as is the case for observations with UVES at the VLT. For the main-sequence star, we obtain SNR per pixel $\sim$ 200 and 100, for the bright and faint cases, respectively. For the subgiant, the values are SNR $\sim$ 170 and 85, for the bright and faint cases, respectively. The lower SNR of the subgiant, reflects the lower flux that it outputs in the near-UV wavelengths because of its lower $T_\mathrm{eff}$. In addition to these SNR values, we consider a case of higher SNR that simulates the gain obtained by stacking four observations of each star (i.e., a total of four observing hours spent in each target). This option of stacked spectra have SNR = 400, for the main-sequence stars, and SNR = 340, for the subgiant stars. For comparison, the observations of the same stars were simulated using the UVES ETC, assuming similar observing conditions and using the blue arm centered at 346nm with the 1.0 arcsec slit. For the main sequence star, the obtained SNR values are 30 and 13, for the bright and faint case, respectively. For the subgiant, the values are SNR = 18 and 7, for the bright and faint case, respectively.

Gaussian noise equivalent to these SNR levels were added to the spectra using the task {\sf mknoise} in IRAF\footnote{Used within AstroConda, which is maintained by the Space Telescope Science Institute (STScI). See \url{https://astroconda.readthedocs.io/en/latest/index.html}.}.  In total, four realizations of the noise were created, so that an indication of the random uncertainties affecting the analysis could be obtained. Four noise simulations of 80 spectra in three different SNR levels results in a total of 960 spectra for our analysis.

\begin{figure}
    \centering
    \includegraphics[height=6cm]{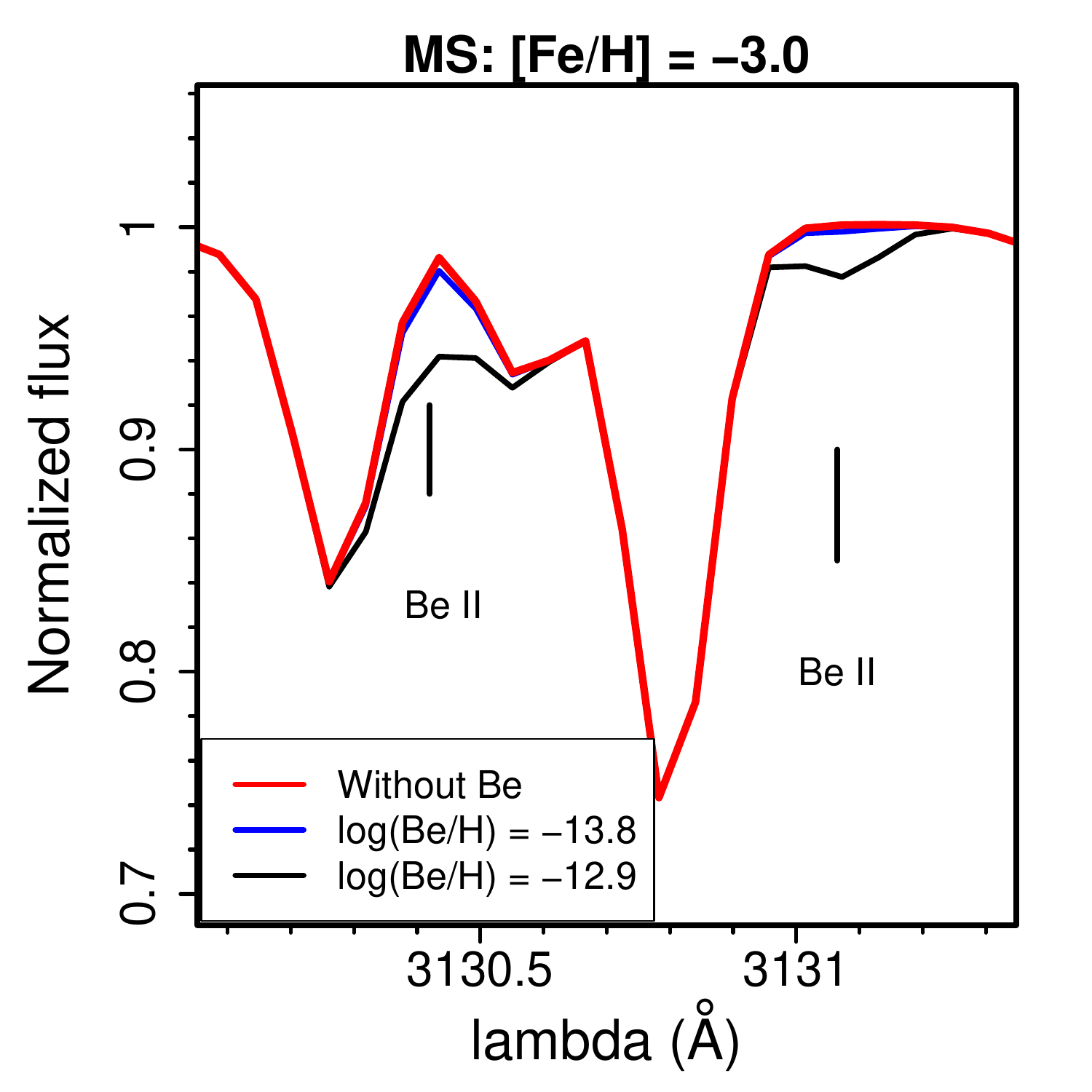}
    \includegraphics[height=6cm]{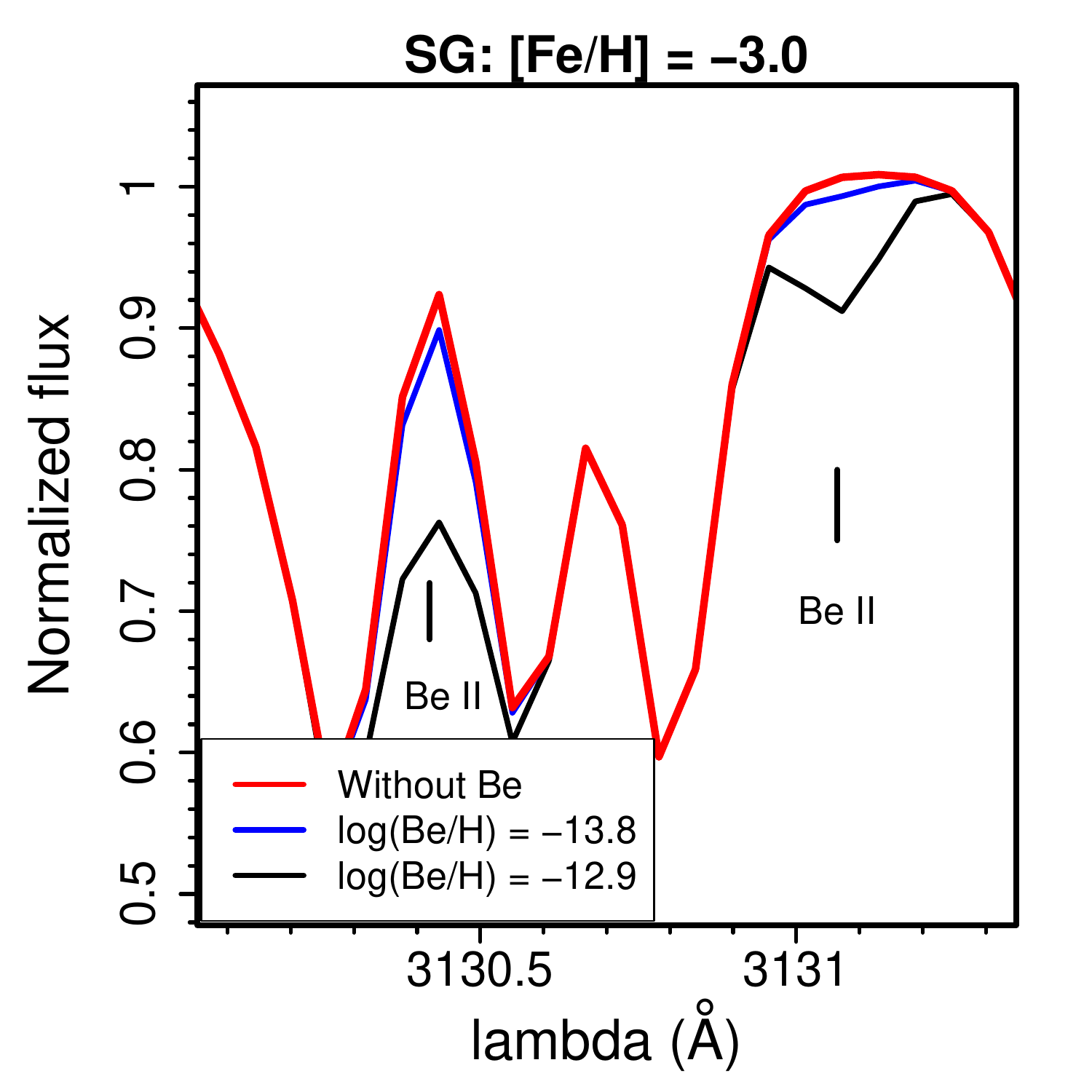}
    \includegraphics[height=6cm]{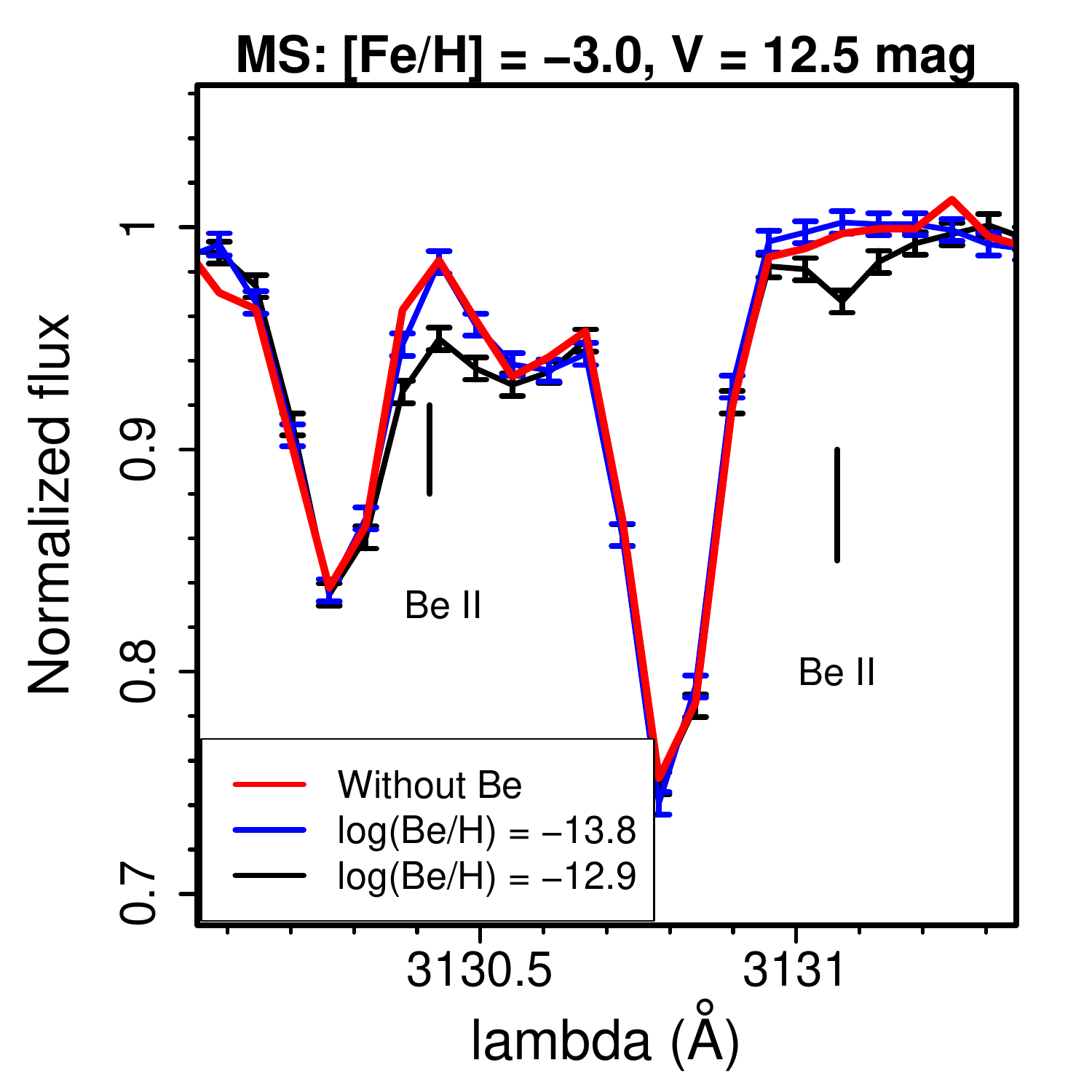}
    \includegraphics[height=6cm]{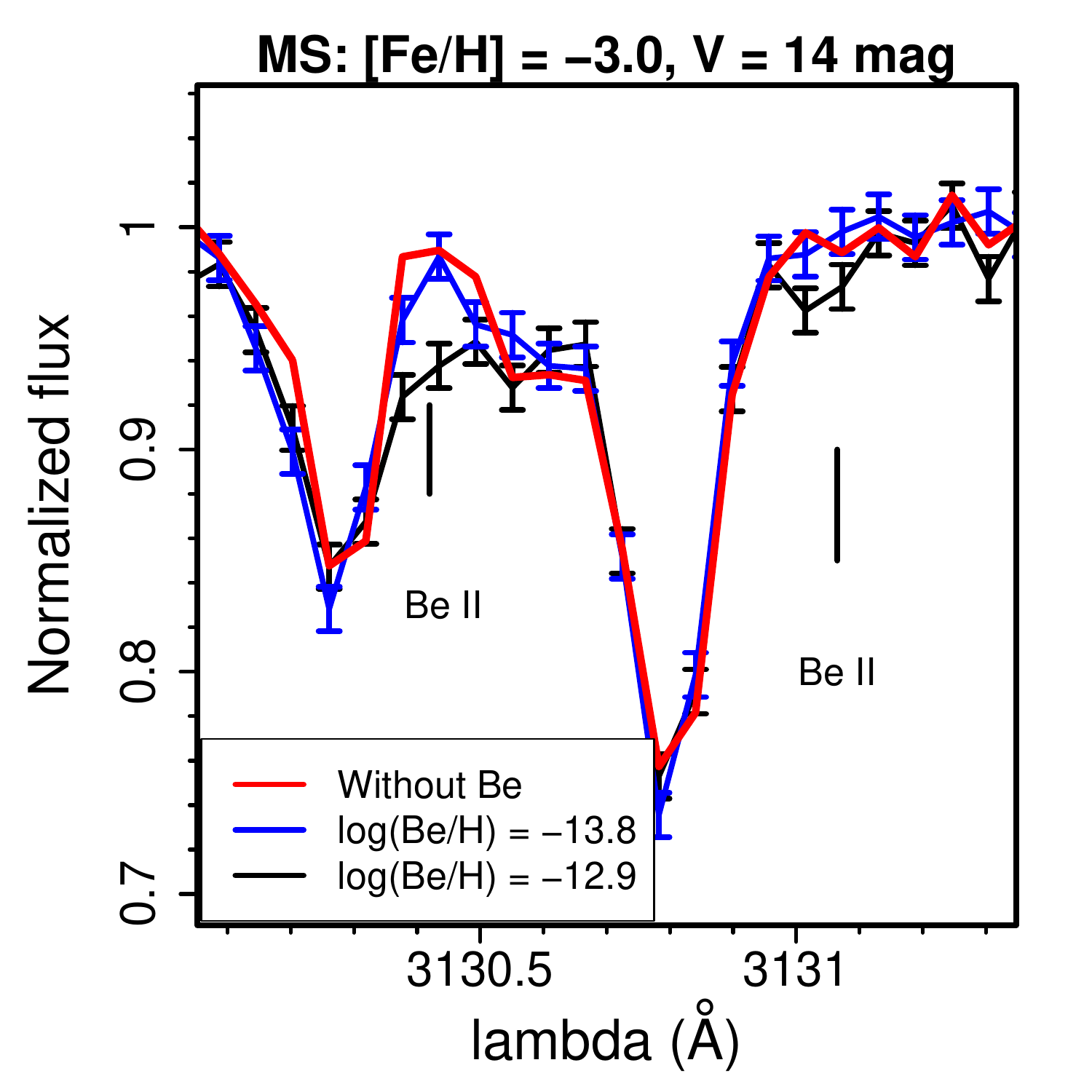}
    \includegraphics[height=6cm]{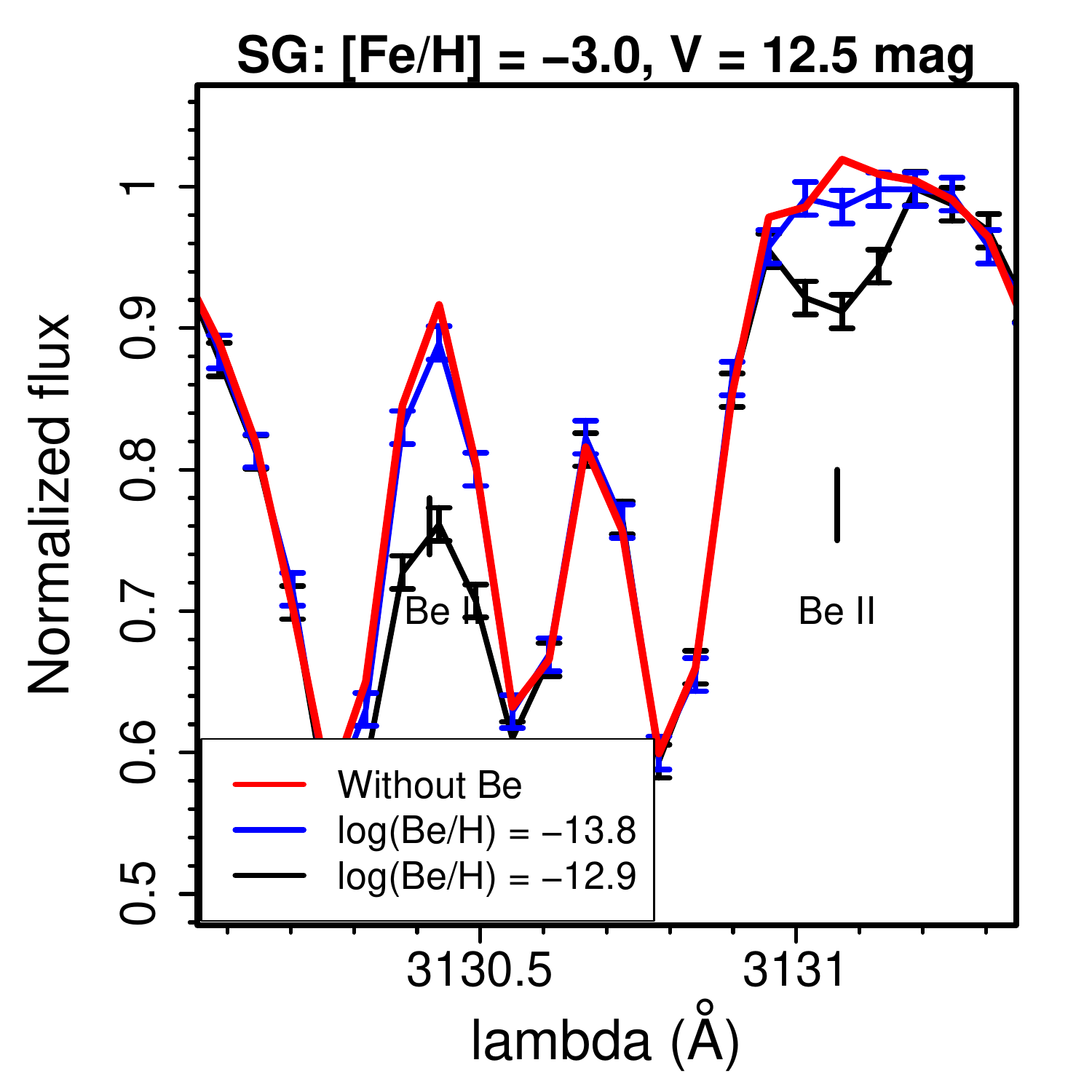}
    \includegraphics[height=6cm]{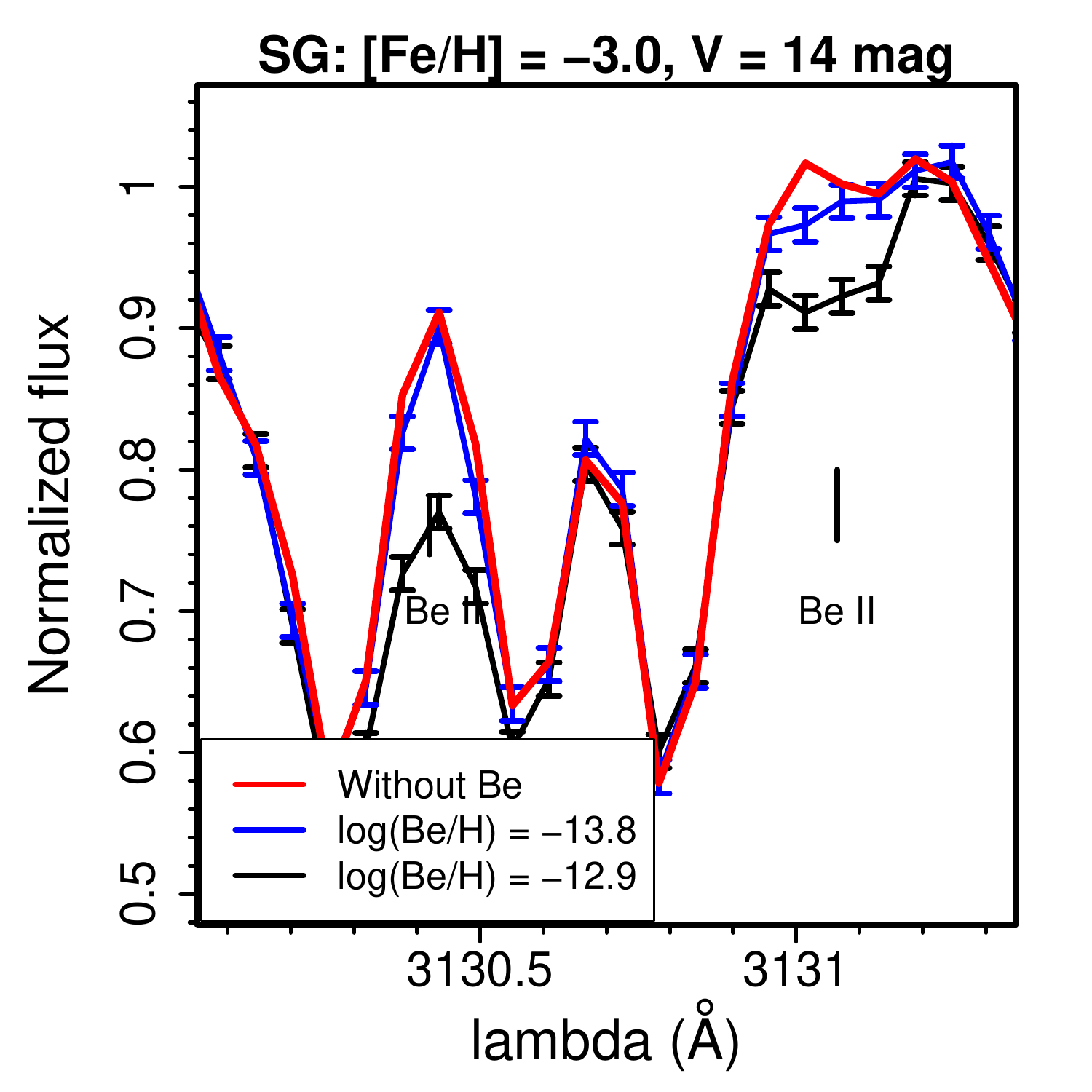}
    \caption{Comparison of the stellar spectra, with and without noise, for three different beryllium abundances. Note the different y-axis scale for the panels showing the subgiant stars. The Be abundances are given in the insets of each panel. \textit{Top row:} Normalized spectra for the main-sequence star (left) and the subgiant star (right) with [Fe/H] = $-$3.0, without noise. \textit{Middle row:} Normalized spectra for the bright (left) and faint (right) main-sequence star with [Fe/H] = $-$3.0. The length of the error bars indicate the $\pm$1$\sigma$ error for SNR = 200 and 100 for the bright and faint cases, respectively. \textit{Bottom row:} Normalized spectra for a bright (left) and faint (right) subgiant star with [Fe/H] = $-$3.0 (SNR = 170 and 85, for the bright and fain stars, respectively).}
    \label{fig:spectra}
\end{figure}

\section{Detecting the weak Be lines}\label{sec:limits}

The magnitude of the challenge that we want to address can be appreciated in Fig.\ \ref{fig:spectra}. For the main-sequence star, the Be lines are very weak and hard to be appropriately measured even in the case without noise. It is clear that there is a change in the detection limit with SNR, when moving from the bright to the faint case. For the subgiant star, which is cooler and has lower $\log~g$ value, the Be lines are stronger when compared to the main sequence star, even if the abundance values are the same. The detection of the lines is easier for this type of star. Besides the effect of the SNR, it is important to note from these plots that these weak Be lines are spread over only three to maybe five pixels in the CUBES spectra.

To determine if the Be lines in a given noised spectrum can be detected, we proceeded as follows. Each noised observed spectrum of a given star is compared to the 20 synthetic spectra, without noise, of the same star (where the Be abundances vary by 0.05 dex with one spectrum computed without Be). The first step is to search for the synthetic spectrum that best fits the observations. Here, we simply compute the coefficient of determination, $R^2$, which involves the ratio between the residual sum of squares and the total sum of squares:

\begin{equation}
    R^2 = 1 - \frac{\sum^n_{i=1}\,(O_i-M_i)^2}{\sum^n_{i=1}\,(O_i-\overline{O})^2},
    \label{eq:r2}
\end{equation}

where $n$ is the total number of pixels used in the fitting, $O_i$ is the normalized flux of the simulated observation in pixel $i$, $M_i$ is the normalized flux of the model synthetic spectrum in pixel $i$, and $\overline{O}$ is the mean of all $O_i$ values. We apply this equation in a region including 5 pixels around the center of each Be line. A good model that fits the observations will have $R^2$ approaching one (as the residual sum of squares tends to zero). Model selection using $R^2$ is usually not advised, as models can be built with a large number of free parameters to provide an ever better fit to the data. In our case, however, the only free parameter is the Be abundance and we simply want to find the model that better explains the simulated data. In this procedure, each Be line is analyzed separately. 

After finding the best fitting model, we still test if the Be line can be considered detected or if an upper limit should be reported instead. This is done through the comparison of two distributions of normalized residuals. One distribution from the simulated observation with respect to the best-fitting synthetic spectrum, the other from the simulated observation with respect to the spectrum without Be. In both cases, the residuals were normalized to the error expected from the SNR. These two distributions are then compared using a two-sample Kolmogorov-Smirnov test. The null hypothesis is that there is no difference between the two distributions. If the null hypothesis can not be rejected, then the spectrum without Be gives a fit that is statistically as good as the one with the selected Be abundance. This means that the line could not be clearly detected. If the null hypothesis is rejected, we consider that a detection was obtained. 

\subsection{The subgiant star case}

\begin{figure}
    \centering
    \includegraphics[width=0.75\textwidth]{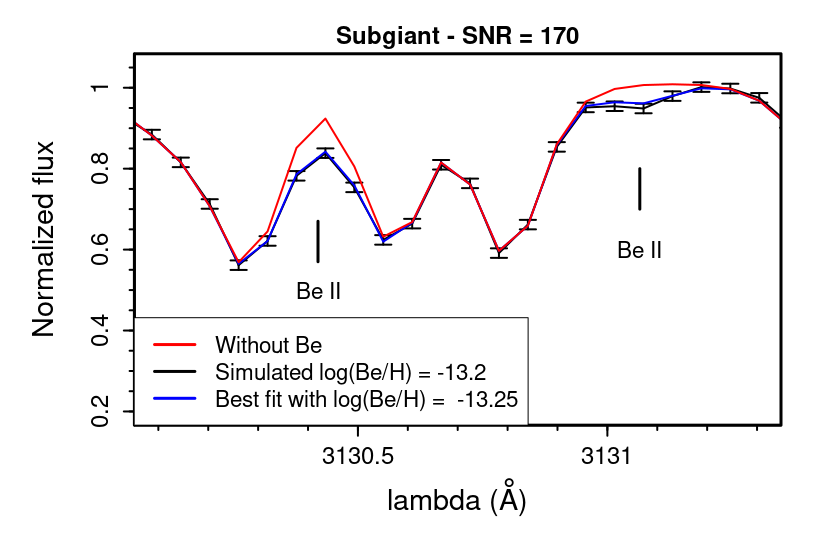}
    \includegraphics[width=0.45\textwidth]{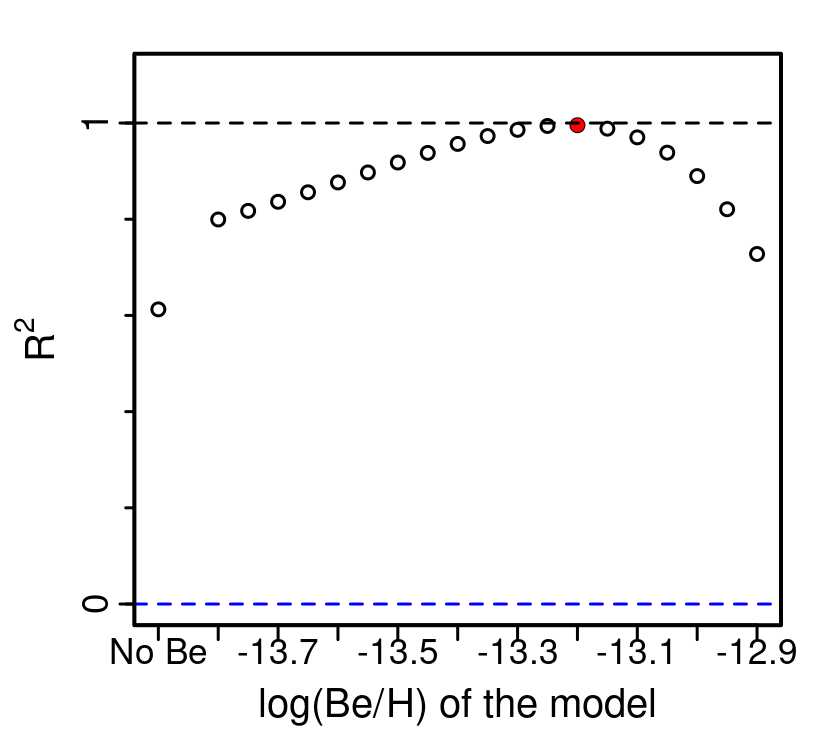}
    \includegraphics[width=0.45\textwidth]{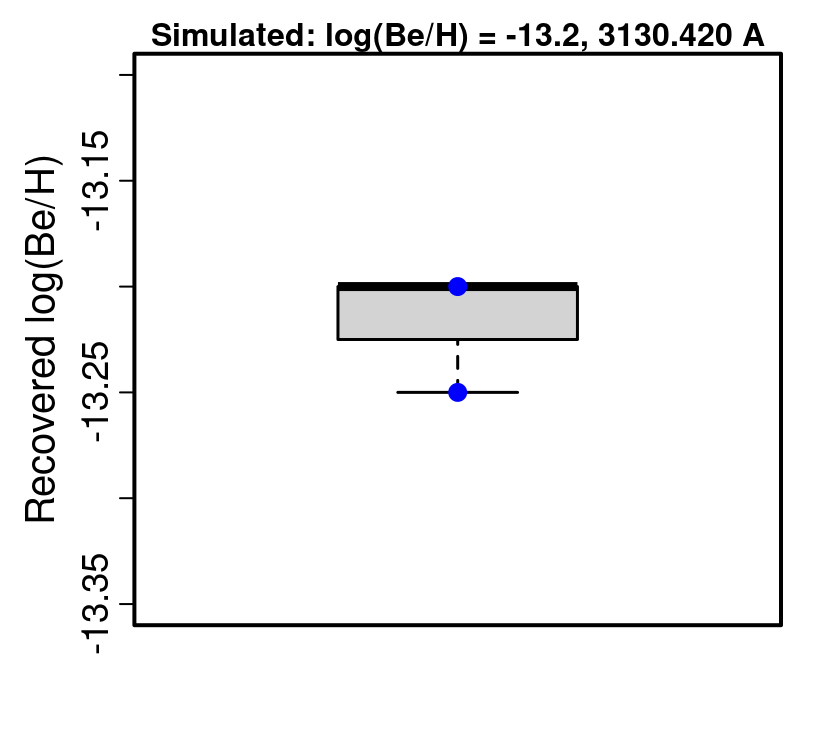}
    \caption{Fitting of the four simulated observations of a bright ($V$ = 12.5 mag) subgiant star with [Fe/H] = -3.0 and $\log(\mathrm{Be/H})$ = $-$13.2, using five pixels around the 3130 \AA\ line. {\bf Top:} One selected fit that returns $\log$(Be/H) = $-$13.25, slightly lower than the input abundance of the observation. {\bf Bottom left:} The $R^2$ values obtained when fitting the Be line with spectra computed with a range of abundances. Note that the spectrum without Be returns a worse fit than all other options that include the Be lines. {\bf Bottom right:} Boxplot of the best fitting abundances in each of the four simulated observations.}
    \label{fig:bright.sg.3130.m3.be13p2}
\end{figure}

\begin{figure}
    \centering
    \includegraphics[width=0.75\textwidth]{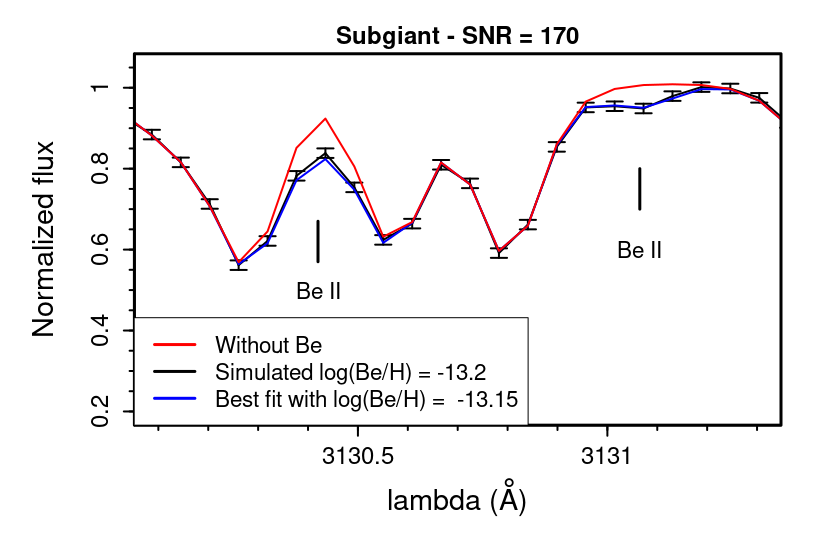}
    \includegraphics[width=0.45\textwidth]{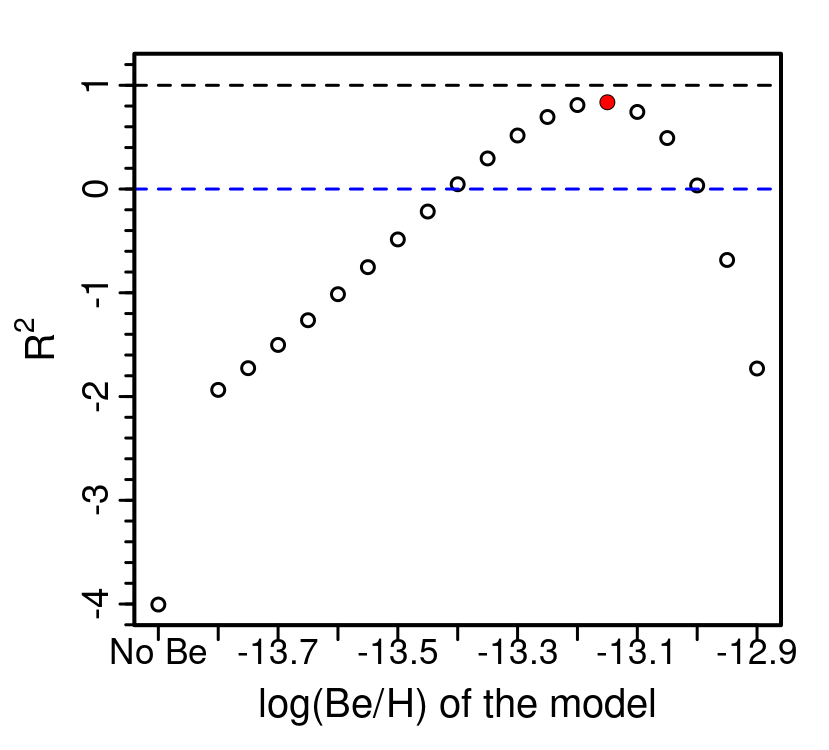}
    \includegraphics[width=0.45\textwidth]{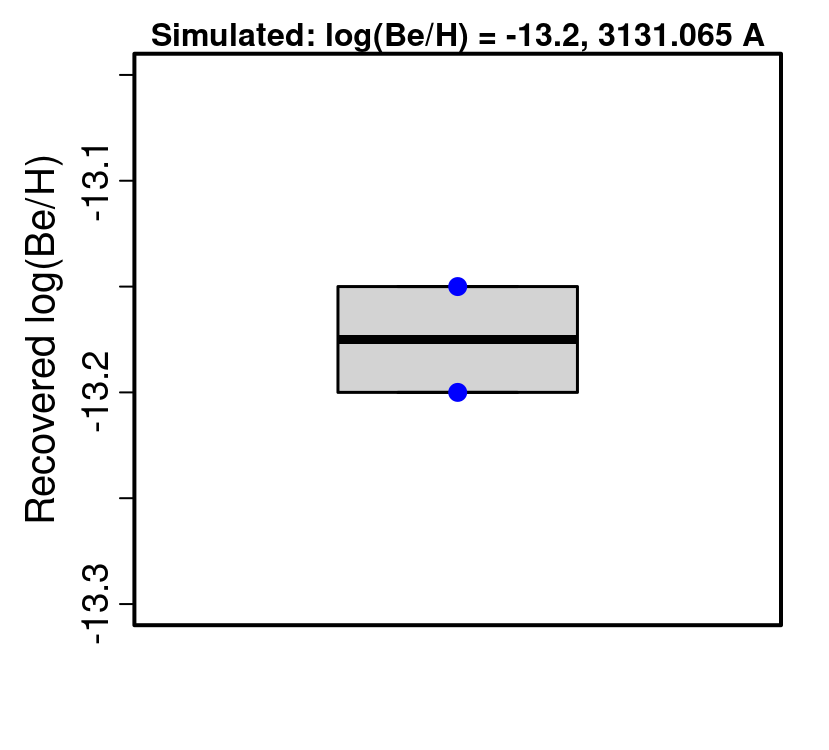}
    \caption{Results of fitting five pixels around the 3131 \AA\ line in the same observations shown in Fig.\ \ref{fig:bright.sg.3130.m3.be13p2}.  {\bf Top:} One selected fit that returns $\log$(Be/H) = $-$13.15 while the input abundance of the observation was $-$13.20. {\bf Bottom left:} The $R^2$ values obtained when fitting the line with spectra of a range of abundances. The spectrum without Be again returns a worse fit than all other options. {\bf Bottom right:} Boxplot of the best fitting abundances in each of the four simulated observations.}
    \label{fig:bright.sg.3131.m3.be13p2}
\end{figure}

As an example, we start looking at the four simulated one hour observations of a bright ($V$ = 12.5 mag) subgiant star with [Fe/H] = -3.0 and $\log(\mathrm{Be/H})$ = $-$13.2. This value of Be abundance was the detection limit for the main-sequence stars observed with UVES analyzed in \citet{2021A&A...646A..70S}. In Figs.~\ref{fig:bright.sg.3130.m3.be13p2} and \ref{fig:bright.sg.3131.m3.be13p2} we show examples of fitting made to the 3130 \AA\ and 3131 lines, respectively. In all four observations, both lines are detected with best fitting abundance differing by at most 0.05 dex of the input one.

For the 3130 \AA\ line, three realizations of the observed spectrum are fit with Be at the same value of the input abundance, $\log(\mathrm{Be/H})$ = $-$13.2. The fourth observed spectrum is fit with a slightly lower value, $\log(\mathrm{Be/H})$ = $-$13.25 (this last case is shown at top panel of Fig.\ \ref{fig:bright.sg.3130.m3.be13p2}. A boxplot of the four abundances is shown in the bottom right panel.). Looking at the $R^2$ values of the fits, however, we note that even if there is one fit that gives a maximum value close to one, there are other fits, with abundances within $\pm$0.05-0.10 dex of the best fit, that return very similar $R^2$ values (bottom left panel of Fig.\ \ref{fig:bright.sg.3130.m3.be13p2}). This suggests that, in this case, the fitting uncertainty is of the order of $\pm$0.10 dex. In the fitting of all four simulated observations, the Kolmogorov-Smirnov test indicates that the line was detected.

For the 3131 \AA\ line the results are very similar, even if this line is weaker. Two of the noise realizations are fit with Be at the same value of the input abundance, $-$13.2, but two are fit by a slightly higher value, $-$13.15 (one of them is shown at top panel of Fig.\ \ref{fig:bright.sg.3131.m3.be13p2}. The boxplot of the four abundances is shown in the bottom right panel). The $R^2$ values here span a wider range. This happens in part because of the term at the bottom part of Equation \ref{eq:r2}, which involves the average of the observed flux values. For the stronger 3130 line, which is also affected by strong blends, this term is larger which has the effect of reducing the scale of the total $R^2$ variation. The $R^2$ values indicate that abundances within $\pm$0.05 dex of the best value give results of similar quality (bottom left panel of Fig.\ \ref{fig:bright.sg.3131.m3.be13p2}). In the fitting of all four simulated observations, the Kolmogorov-Smirnov test indicates that the line was detected.  

To determine the detection limit, we analyse the other simulated observations in the same way, looking at the ones with progressively lower Be abundance, in steps of 0.05 dex. For the weaker line at 3131 \AA, we reach a quite different situation already when the input abundance of the observations is $\log$(Be/H) = $-$13.25. The best fitting synthetic spectra for the four observations agree with the input within $\pm$0.05, the $R^2$ values indicate a clear peak close to unity around the same values, but in three cases the Kolmogorov-Smirnov test indicates that the line was not detected. Exactly the same happens for $\log$(Be/H) = $-$13.3 and $-$13.35. For the next step, $-$13.4, all four Kolmogorov-Smirnov tests indicate that a detection was not possible. Moreover, the $R^2$ values now fail to approach unit, remaining between 08-0.9 for three cases, and with one case where the maximum value is below zero. While this last abundance level seems clearly not detected, the case for the other values seem more ambiguous. We conclude that the limit for a detection of the 3131\AA~Be line in this case is at an abundance of $\log$(Be/H) = $-$13.25.

\begin{figure}
    \centering
    \includegraphics[width=0.75\textwidth]{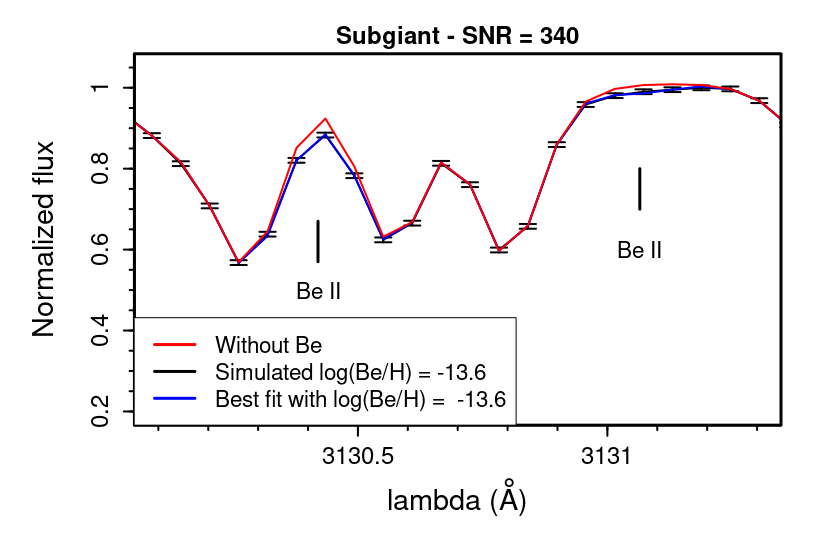}
    \includegraphics[width=0.45\textwidth]{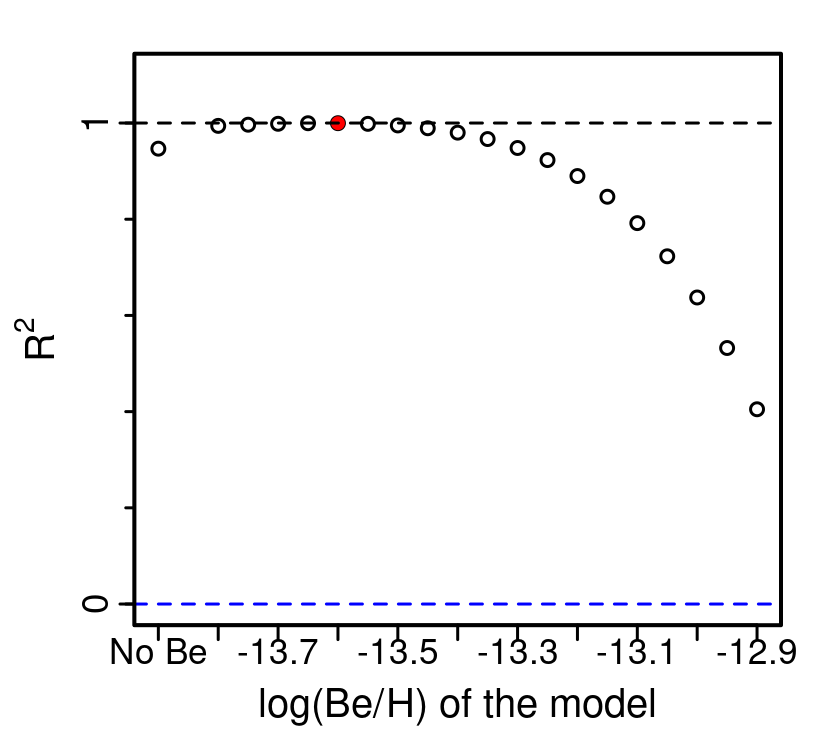}
    \includegraphics[width=0.45\textwidth]{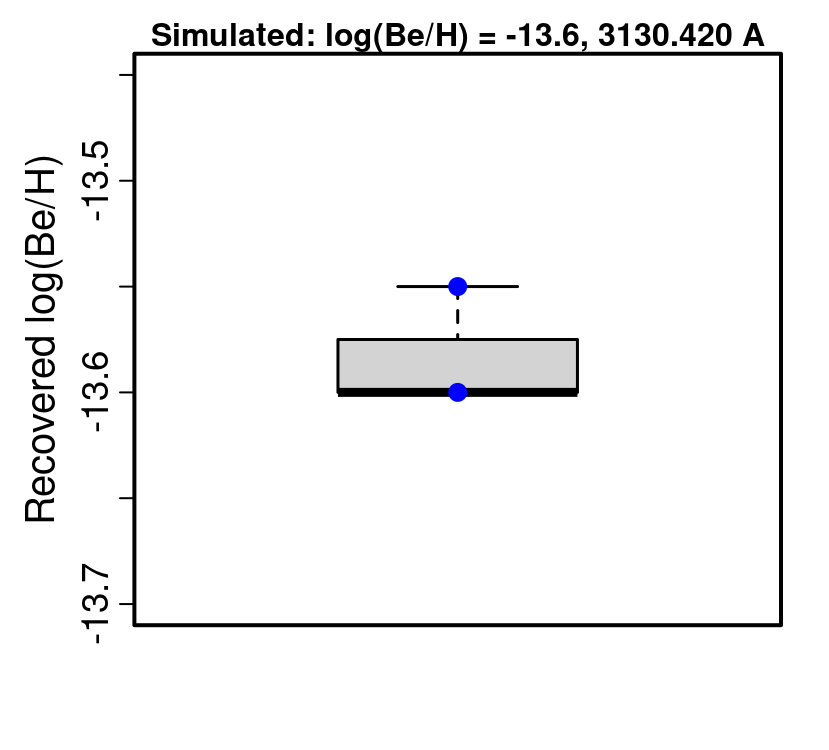}
    \caption{Same as Fig.~\ref{fig:bright.sg.3130.m3.be13p2}, fitting the 3130 \AA~line for an input abundance of $\log(\mathrm{Be/H})$ = $-$13.6. {\bf Top:} One selected best fit with $\log$(Be/H) = $-$13.6, the same value of the input abundance of the observation. {\bf Bottom left:} The $R^2$ values obtained when fitting the Be line with spectra with a range of abundances. Note that the range is different from the previous plots. Several of the fitted spectra provide $R^2$ values above 0.8. {\bf Bottom right:} Boxplot of the best fitting abundances in each of the four simulated observations.}
    \label{fig:bright.sg.3130.m3.be13p5}
\end{figure}

The line at 3130 \AA~is stronger, so visually it seems that it could be detected at lower values of the abundance. However, already at the same value of $\log$(Be/H) = $-$13.25 some ambiguity starts to appear. The best fitting abundances are within $\pm$0.05 dex of the input, the $R^2$ plots indicate that abundances within $\pm$0.10 of the best value provide similar fits, but one of the four Kolmogorov-Smirnov tests indicate that a detection was not possible. Essentially the same happens for observations with $\log$(Be/H) = $-$13.3 and $-$13.35, although now the $R^2$ plots indicate some degeneracy in the fits of maybe $\pm$0.15. Only for abundances below $\log$(Be/H) = $-$13.4, two or more Kolmogorov-Smirnov tests indicate that a detection was not possible. From these considerations, we conclude that the limit for a clear detection of the 3130\AA~Be line is most likely around an abundance of $\log$(Be/H) = $-$13.35, which is the last case before the analysis starts to return mostly dubious fits.

If we increase the SNR of the observations, for a case of four exposures of the bright subgiant (which is also equivalent to 16 exposures of the faint subgiant), the limits improve. For the 3130 \AA~line, most tests indicate that a detection is possible down to $\log$(Be/H) = $-$13.6 (see Fig.\ \ref{fig:bright.sg.3130.m3.be13p5}). At this point, however, the $R^2$ plot shows that there is already no difference between spectra with abundances $\pm$0.15 dex of the best fit. For the weaker Be line, the tests indicate we can detect the line down to $\log$(Be/H) = $-$13.5. Finally, for the case of lower SNR (one single exposure of the faint subgiant), our tests indicate a detection only for the highest level of the abundances that we simulated, $\log$(Be/H) = $-$12.95.

For the case of a subgiant of lower metallicity, [Fe/H] = $-$3.5, we only simulated abundances below $\log$(Be/H) = $-$13.2. At this level, for the cases of SNR = 85, no clear detection seems possible. At SNR = 170, line 3130 \AA~seems to be detectable at lower Be abundances when compared to the case with [Fe/H] = $-$3.0. On the other hand, line 3131 \AA~seems detectable only at a slightly higher Be abundance. For the line at 3130 \AA, this difference might have to do with the reduced influence of the blends surrounding it. For the other line, it seems just part of the statistical fluctuation. In the same way, for the case with SNR = 340, the line at 3130 \AA\, seems to be detectable down to a lower abundance, $\log$(Be/H) = $-$13.7, with little difference among the fits with $\pm$0.15 dex.

\begin{table}[]
    \centering
    \begin{tabular}{cccccc}
       \hline
      $V$ (mag) & t$_{\rm exp}$ & SNR & {[Fe/H]} & Line 3130 \AA~ & Line 3131\AA~ \\
         \hline
         \hline
      14 & 3000s & 85 & $-$3.00 & $\geq$ $-$12.95 & $\geq$ $-$12.95 \\
      14 & 3000s & 85 & $-$3.50 & --  & -- \\
      12.5 & 3000s & 170 & $-$3.00 & $\geq$ $-$13.35 & $\geq$ $-$13.25 \\
      12.5 & 3000s & 170 & $-$3.50 & $\geq$ $-$13.50 & $\geq$ $-$13.20 \\
      12.5 & 4 $\times$ 3000s & 340 & $-$3.00 & $\geq$ $-$13.60 & $\geq$ $-$13.50 \\
      12.5 & 4 $\times$ 3000s & 340 & $-$3.50 &  $\geq$ $-$13.70 & $\geq$ $-$13.45 \\
        \hline
    \end{tabular}
    \caption{Limits of beryllium abundance detection for the subgiant case. Note that the case of SNR = 170 can also be obtained with four 3000s exposures for the faint, $V$ = 14 mag, star. The case with SNR = 340 would require 16 exposures of 3000s for the same faint star.}
    \label{tab:Belimits.subgiant}
\end{table}
Overall, the conclusion here seems to be that, as long as the SNR approaches a value close to 400, with CUBES we can push the detection limit to $\log$(Be/H) = $-$13.6 or $-$13.7, with an uncertainty of the order of $\pm$0.15 dex. The derived detection limits for each level of SNR are given in Table \ref{tab:Belimits.subgiant}.

\subsection{The main-sequence star case}

Finally, let us examine the case for the observations of a main sequence star. For this case, the Be lines are weaker, so we can anticipate detection limits at higher abundances. With SNR = 100, neither of the two lines can be detected, even at the maximum Be abundances that we simulated ($\log$(Be/H) = $-$12.9 for [Fe/H] = $-$3.0 and $\log$(Be/H) = $-$13.2 for [Fe/H] = $-$3.5).

Even increasing the SNR to 200, the line at 3131 \AA\ remains not detectable. It becomes marginally possible to detect this line only with SNR = 400 and only at the highest abundances simulated for the [Fe/H] = $-$3.0 case. For $\log$(Be/H) = $-$12.9, two of the fits pass the Kolmogorov-Smirnov test and the $R^2$ values are above 0.8. 

\begin{figure}
    \centering
    \includegraphics[width=0.75\textwidth]{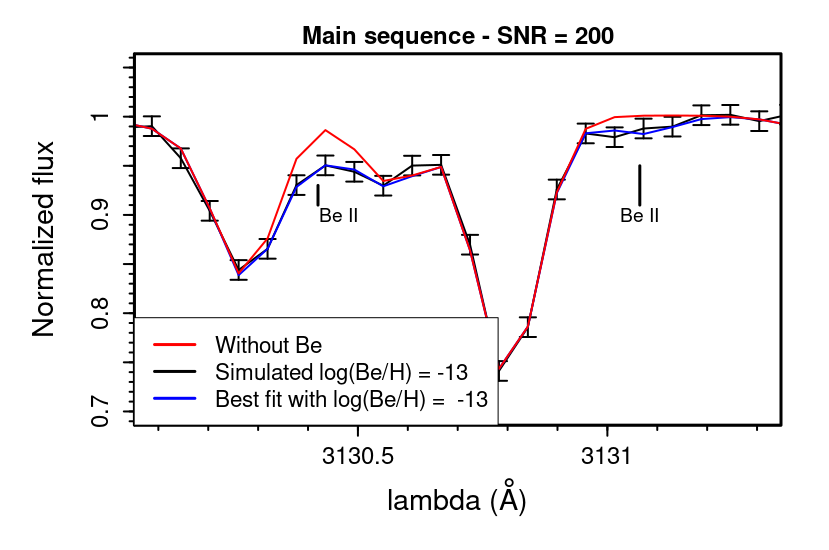}
    \includegraphics[width=0.45\textwidth]{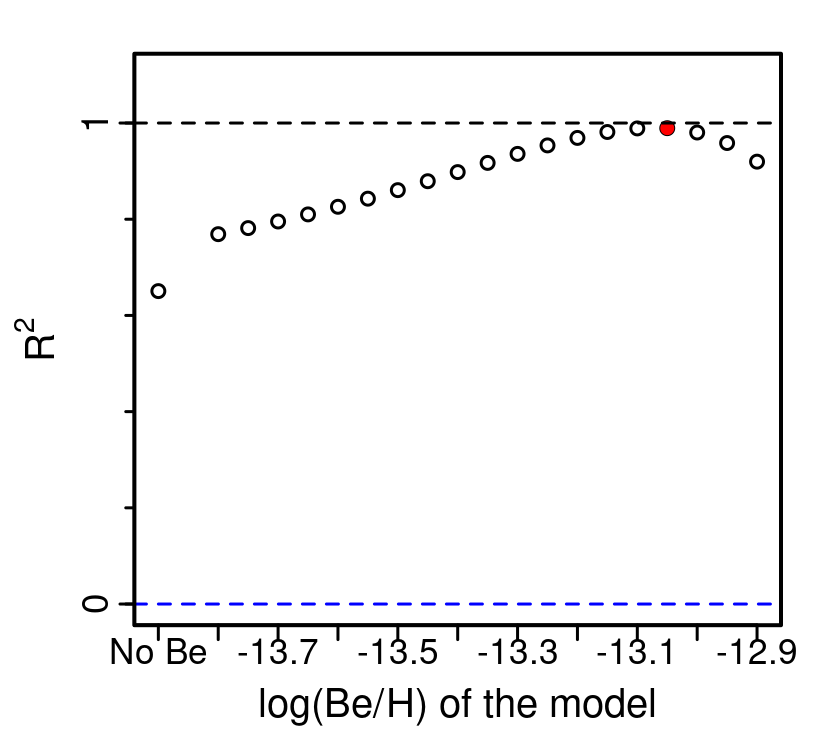}
    \includegraphics[width=0.45\textwidth]{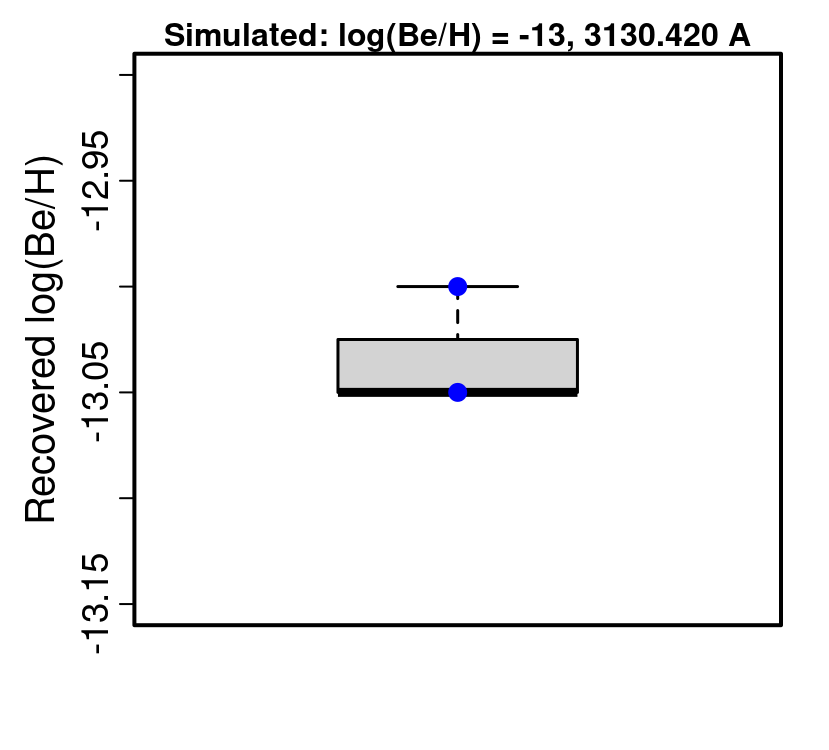}
    \caption{Same as Fig.~\ref{fig:bright.sg.3130.m3.be13p2}, but fitting the 3130 \AA\ line in the case of a bright main-sequence star with [Fe/H] = $-$3.0 and input abundance of $\log(\mathrm{Be/H})$ = $-$13.0, observed for 3000s. {\bf Top:} One selected best fit with $\log$(Be/H) = $-$13.0, the same value of the input abundance of the observation. {\bf Bottom left:} The $R^2$ values obtained when fitting the Be line with spectra with a range of abundances. {\bf Bottom right:} Boxplot of the best fitting abundances in each of the four simulated observations.}
    \label{fig:bright.ms.3130.m3.be13p0}
\end{figure}

It seems that it is only the stronger line at 3130 \AA\ that stands a chance of really being detected. At the same SNR = 200, we determined that the limit is at $\log$(Be/H) = $-$13.0 (Fig.\ \ref{fig:bright.ms.3130.m3.be13p0}). The abundances of the best fitted spectra are within $\pm$0.05 dex of the input value, there is a clear peak in the $R^2$ values, and the Kolmogorov-Smirnov test is still consistent with a detection. For lower abundances in the observed spectra, the fits are still good, both visually and in the $R^2$ plot. However, the Kolmogorov-Smirnov tests already indicate that the noise level precludes a clear detection. For the main-sequence star of lower metallicity, [Fe/H] = -3.5, our simulation with the highest Be abundance, $\log$(Be/H) = $-$13.2, is not detectable.

\begin{figure}
    \centering
    \includegraphics[width=0.75\textwidth]{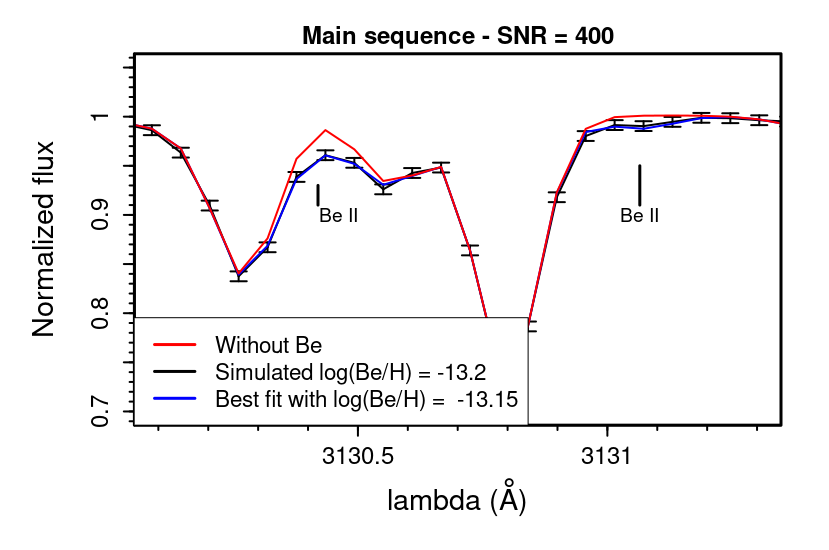}
    \includegraphics[width=0.45\textwidth]{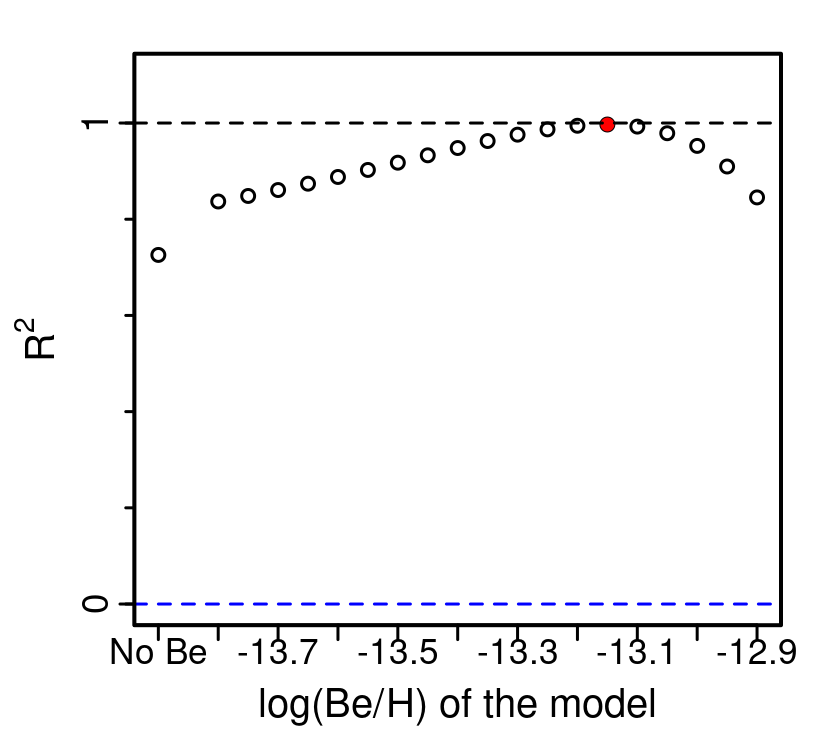}
    \includegraphics[width=0.45\textwidth]{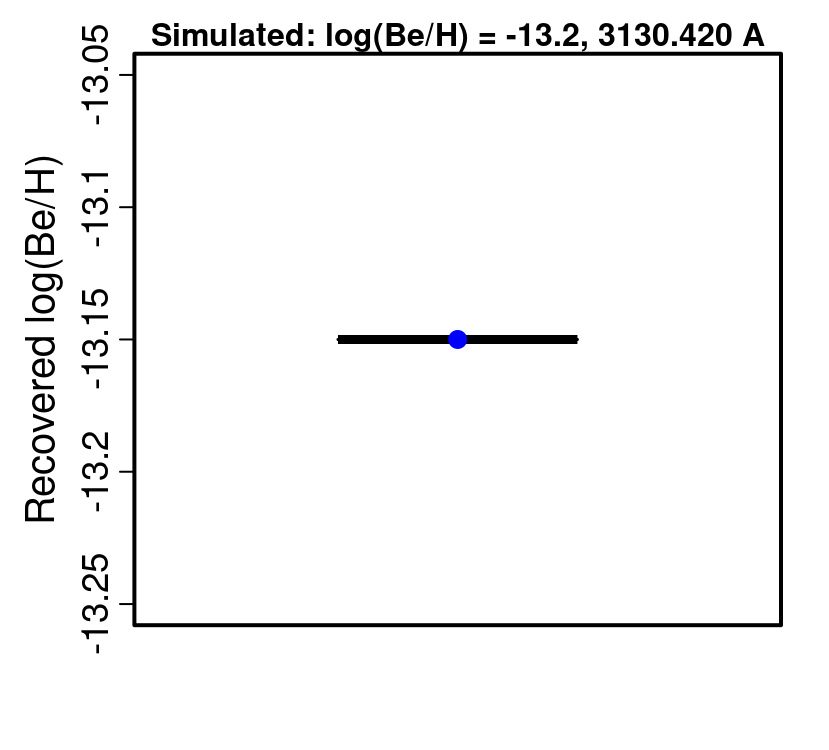}
    \caption{Same as Fig.~\ref{fig:bright.ms.3130.m3.be13p0}, but for the case where SNR = 400. {\bf Top:} One selected best fit with $\log$(Be/H) = $-$13.15 for the observation that was simulated with input abundance of $\log$(Be/H) = $-$13.20. {\bf Bottom left:} The $R^2$ values obtained when fitting the Be line with spectra with a range of abundances. {\bf Bottom right:} Boxplot of the best fitting abundances in each of the four simulated observations.}
    \label{fig:bright.ms.3130.m3.be13p2}
\end{figure}

\begin{figure}
    \centering
    \includegraphics[width=0.75\textwidth]{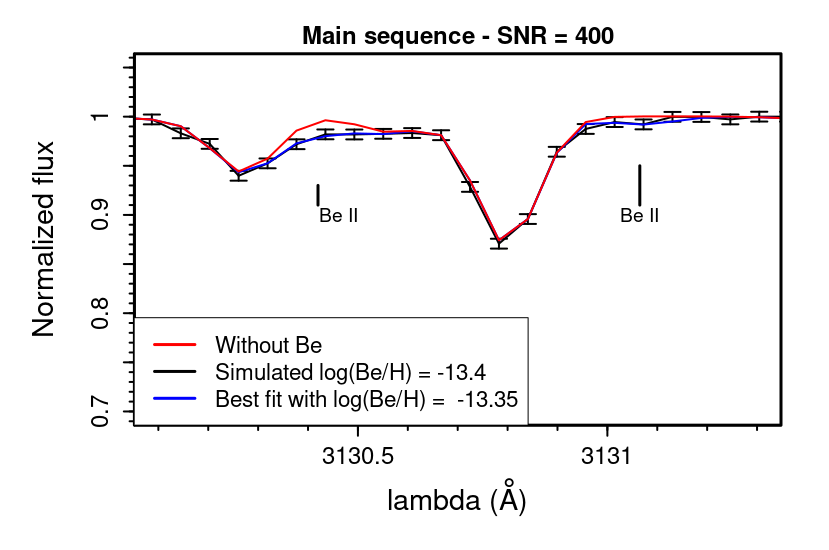}
    \includegraphics[width=0.45\textwidth]{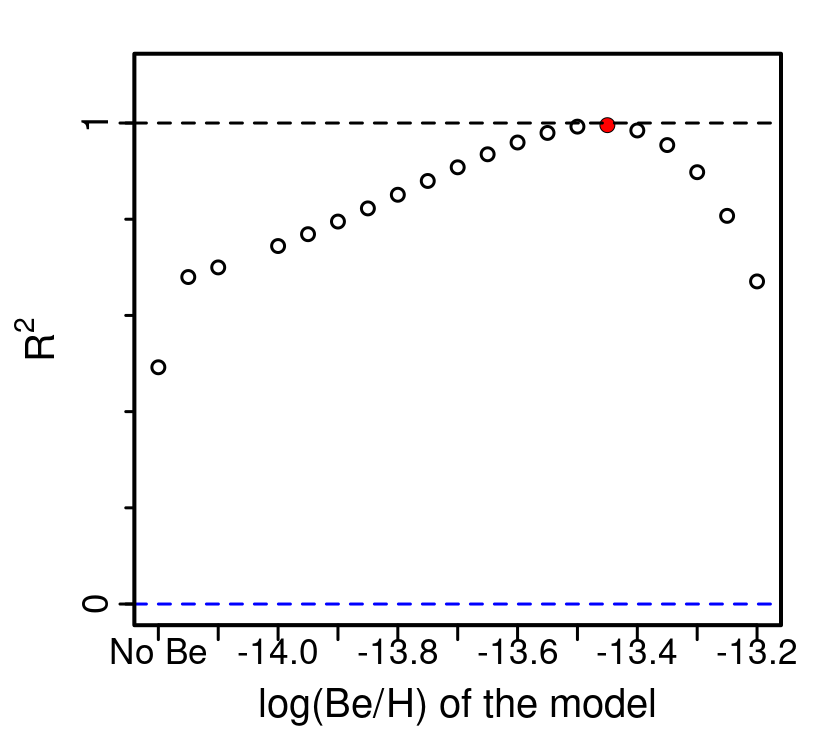}
    \includegraphics[width=0.45\textwidth]{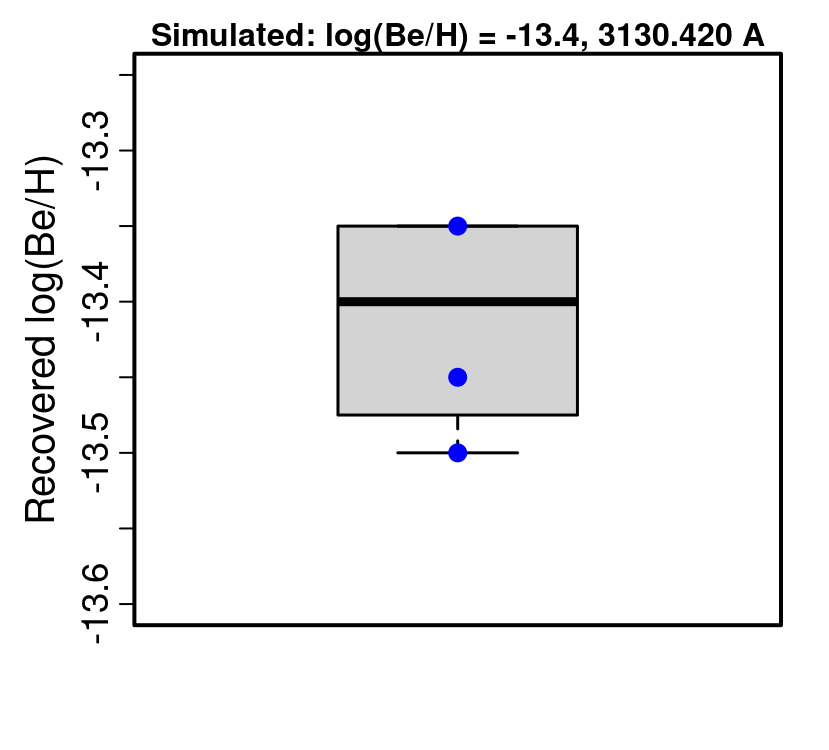}
    \caption{Same as Fig.~\ref{fig:bright.ms.3130.m3.be13p2}, but for the case of star with [Fe/H] = $-$3.5. {\bf Top:} One selected best fit with $\log$(Be/H) = $-$13.35 for the observation that was simulated with input abundance of $\log$(Be/H) = $-$13.40. {\bf Bottom left:} The $R^2$ values obtained when fitting the Be line with spectra with a range of abundances. {\bf Bottom right:} Boxplot of the best fitting abundances in each of the four simulated observations.}
    \label{fig:bright.ms.3130.m3.be13p4}
\end{figure}

The final case is where the stacked observations of the main-sequence star result in SNR = 400. As was the case for the subgiant star, the detection limit seems to depend on how strong are the surrounding blends. For a star with [Fe/H] = $-$3.00, the lower abundance that we can detect is $\log$(Be/H) = $-$13.2 (Fig.\ \ref{fig:bright.ms.3130.m3.be13p2}). In this case, the $R^2$ plot indicates that synthetic spectra with abundances within $\pm$0.10 dex return fits of similar quality. For a star with [Fe/H] = $-$3.50, the lower abundance we can detect is smaller, $\log$(Be/H) = $-$13.4 (Fig.\ \ref{fig:bright.ms.3130.m3.be13p4}). In this case, the boxplot shows that the best fits were obtained with abundances within $\pm$0.15 dex of the input value. The $R^2$ plots indicates that synthetic spectra with abundances within $\pm$0.10 dex return fits of similar quality. For the main-sequence star case, the detection limits for each level of SNR are given in Table \ref{tab:Belimits.ms}.

\begin{table}[]
    \centering
    \begin{tabular}{cccccc}
       \hline
      $V$ (mag) & t$_{\rm exp}$ & SNR & {[Fe/H]} & Line 3130 \AA~ & Line 3131\AA~ \\
         \hline
         \hline
      14 & 3000s & 100 & $-$3.00 & -- & -- \\
      14 & 3000s & 100 & $-$3.50 & --  & -- \\
      12.5 & 3000s & 200 & $-$3.00 & $\geq$ $-$13.0 & -- \\
      12.5 & 3000s & 200 & $-$3.50 & -- & -- \\
      12.5 & 4 $\times$ 3000s & 400 & $-$3.00 & $\geq$ $-$13.2 & $\geq$ $-$12.90 \\
      12.5 & 4 $\times$ 3000s & 400 & $-$3.50 & $\geq$ $-$13.4 & -- \\
        \hline
    \end{tabular}
    \caption{Limits of beryllium abundance detection for the mains-sequence star case. Note that the case of SNR = 200 can also be obtained with four 3000s exposures for the faint star. The case with SNR = 400 would require 16 exposures of 3000s for the faint star.}
    \label{tab:Belimits.ms}
\end{table}

\section{Conclusions}\label{sec:conclusions}

In the context of the phase A study of CUBES, we investigated the limits for detecting weak Be lines in the spectra of extremely metal-poor stars ([Fe/H] $\leq$ $-$3.0). The current design of CUBES, described in Zanutta et al. (this Special Issue), shows that it should be possible to obtain spectra with R $\sim$ 23\,000 and a sampling of 2.35 pixels in the region of the Be lines. 

For the simulations, we estimated the SNR that can be obtained using the CUBES ETC (Genoni et al., this Special Issue). We considered a standard service mode observing block of one hour with ten minutes of overhead and 50 minutes of exposure time. To compute the spectra, we adopted the line list described in the companion paper of Giribaldi et al. (this Special Issue) with the inclusion of the HFS splitting of the Be lines. We simulated the spectra of two stars, one subgiant with parameters similar to the star analysed by \citet{2019A&A...624A..44S} and one main-sequence star with parameters similar to the sample analysed in \citet{2021A&A...646A..70S}.

\begin{figure}
    \centering
    \includegraphics[width=0.75\textwidth]{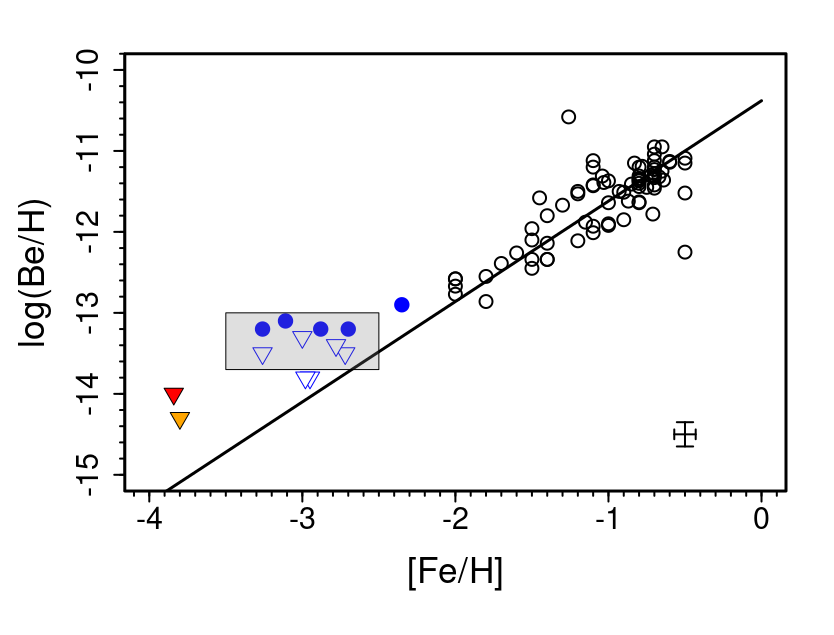}
    \caption{Relation between the Be abundance and metallicity. The upper limits for stars 2MASS J18082002-5104378 \citep{2019A&A...624A..44S} and BD+44 493 \citep{2014ApJ...790...34P} are shown as upside-down red and orange triangles, respectively. The blue circles and blue triangles (for upper limits) are stars analyzed in  \citet{2021A&A...646A..70S}. The black circles are stars from \citet{2009A&A...499..103S}. The solid line is a fit to this last data set. The gray box indicates the parameter space region with $\log$(Be/H) $\geq$ $-$13.7 for stars with $-$2.5 $\geq$ [Fe/H] $\geq$ $-$3.5 where CUBES can potentially be used to detect Be abundances in extremely metal-poor stars and improve the current statistics.}
    \label{fig:be.region}
\end{figure}

The simulations indicate that, as long as a SNR of about 400 can be reached, it should be possible to detect the stronger Be line (at 3130 \AA) with abundances of $\log$(Be/H) $\geq$ $-$13.4 dex and of $\log$(Be/H) $\geq$ $-$13.7 dex, for the main-sequence and subgiant cases, respectively. The typical uncertainty of the fitting only can be of the order of $\pm$0.10-0.15 dex. The SNR $\sim$ 400 can be reached in 4 hours for a star of $V$ = 12.5 mag but would still require about 16 hours for stars $V$ $\sim$ 14 mag. In Fig.\ \ref{fig:be.region} we indicate the region where CUBES can potentially detect Be abundances in comparison to the results from \citet{2021A&A...646A..70S}.

A search in the PASTEL catalog \citep{2016A&A...591A.118S} and the SAGA database \citep{2008PASJ...60.1159S} returned 13 known stars with $V$ $\leq$ 14 mag, $\log~g$ $\geq$ 3.4, and [Fe/H] $\leq$ $-$3.0, that could be interesting targets for a Be survey in extremely metal-poor stars (meaning that Be abundances have not been determined for them before). The number of targets can grow to $\sim$ 21 known objects, if we can push the magnitude limit down to $V$ $\sim$ 14.3 mag. Populating the region with [Fe/H] $\leq$ $-$3.00 of the $\log$(Be/H) vs.\ [Fe/H] diagram can help in understanding how inhomogeneous were the early stages of the Galactic chemical enrichment \citep[see][for a discussion]{2021A&A...646A..70S}. Our results suggest that the lower resolution of the CUBES spectra limits its capability to improve the very low upper limits around or below $\log$(Be/H) $\sim$ $-$14, already existing for a few stars in the literature \citep[e.g.,][]{2014ApJ...790...34P,2019A&A...624A..44S}. The real power of CUBES will be in increasing the magnitude limit of the sample that can be potentially analysed down to $\log$(Be/H) $\geq$ $-$13.7.

One should note, however, that our simulations assume a very idealized case for the analysis. We know a priory, and know perfectly well, the values for the atmospheric parameters of the stars, for the abundances of all other elements in the spectrum, and for the total broadening affecting the spectra. In a real case, all these quantities are unknowns that will be estimated with some uncertainty. Those uncertainties will affect the final result of the analysis. Moreover, the only source of noise affecting our simulated observations is one that is perfectly Gaussian. Finally, deviations from the local thermodynamic equilibrium in the formation of the Be lines were not taken into account. Recent results, however, suggest that such effects might be important in metal-poor stars \citep{2022A&A...657L..11K}. We plan to improve the simulations, addressing some of these points, in the next phases of the development of CUBES. It is clear that detecting the weak Be lines will be a challenging endeavour, but it is also clear that this science case can only be addressed with high-efficiency new spectrographs like CUBES.

%
 \section*{Conflict of interest}
 The authors declare that they have no conflict of interest.

 \section*{Data availability}
The line list used to compute our synthetic spectra is available in Github, \url{https://github.com/RGiribaldi/Master-line-list-for-spectral-synthesis-with-Turbospectrum}. The other datasets generated during and/or analysed during the current study are available from the corresponding author on reasonable request. 
 
\bibliographystyle{spbasic}      
\bibliography{smiljanic_cubes}   

\begin{thebibliography}{76}
\providecommand{\natexlab}[1]{#1}
\providecommand{\url}[1]{{#1}}
\providecommand{\urlprefix}{URL }
\expandafter\ifx\csname urlstyle\endcsname\relax
  \providecommand{\doi}[1]{DOI~\discretionary{}{}{}#1}\else
  \providecommand{\doi}{DOI~\discretionary{}{}{}\begingroup
  \urlstyle{rm}\Url}\fi
\providecommand{\eprint}[2][]{\url{#2}}

\bibitem[{{Anstee} and {O'Mara}(1991)}]{1991MNRAS.253..549A}
{Anstee} SD, {O'Mara} BJ (1991) {An investigation of Brueckner's theory of line
  broadening with application to the sodium D lines}. \mnras 253:549--560,
  \doi{10.1093/mnras/253.3.549}

\bibitem[{{Barbuy} et~al.(2014){Barbuy}, {Bawden Macanhan}, {Bristow},
  {Castilho}, {Dekker}, {Delabre}, {Diaz}, {Gneiding}, {Kerber}, {Kuntschner},
  {La Mura}, {Maciel}, {Mel{\'e}ndez}, {Pasquini}, {Pereira}, {Petitjean},
  {Reiss}, {Siqueira-Mello}, {Smiljanic}, and {Vernet}}]{2014Ap&SS.354..191B}
{Barbuy} B, {Bawden Macanhan} V, {Bristow} P, {Castilho} B, {Dekker} H,
  {Delabre} B, {Diaz} M, {Gneiding} C, {Kerber} F, {Kuntschner} H, {La Mura} G,
  {Maciel} W, {Mel{\'e}ndez} J, {Pasquini} L, {Pereira} CB, {Petitjean} P,
  {Reiss} R, {Siqueira-Mello} C, {Smiljanic} R, {Vernet} J (2014) {CUBES:
  cassegrain U-band Brazil-ESO spectrograph}. \apss 354(1):191--204,
  \doi{10.1007/s10509-014-2039-z}

\bibitem[{{Barklem} and {O'Mara}(1998)}]{1998MNRAS.300..863B}
{Barklem} PS, {O'Mara} BJ (1998) {The broadening of strong lines of Ca\^+,
  Mg\^+ and Ba\^+ by collisions with neutral hydrogen atoms}. \mnras
  300(3):863--871, \doi{10.1046/j.1365-8711.1998.01942.x}

\bibitem[{{Barklem} and {O'Mara}(2000)}]{2000MNRAS.311..535B}
{Barklem} PS, {O'Mara} BJ (2000) {Broadening of lines of Beii, Srii and Baii by
  collisions with hydrogen atoms and the solar abundance of strontium}. \mnras
  311(3):535--540, \doi{10.1046/j.1365-8711.2000.03090.x}

\bibitem[{{Beers} et~al.(2000){Beers}, {Suzuki}, and
  {Yoshii}}]{2000IAUS..198..425B}
{Beers} TC, {Suzuki} TK, {Yoshii} Y (2000) {The Light Elements Be and B as
  Stellar Chronometers in the Early Galaxy}. In: {da Silva} L, {de Medeiros} R,
  {Spite} M (eds) The Light Elements and their Evolution, vol 198, p 425,
  \eprint{astro-ph/0002056}

\bibitem[{{Belokurov} et~al.(2018){Belokurov}, {Erkal}, {Evans}, {Koposov}, and
  {Deason}}]{2018MNRAS.478..611B}
{Belokurov} V, {Erkal} D, {Evans} NW, {Koposov} SE, {Deason} AJ (2018)
  {Co-formation of the disc and the stellar halo}. \mnras 478(1):611--619,
  \doi{10.1093/mnras/sty982}, \eprint{1802.03414}

\bibitem[{{Boesgaard} and {Novicki}(2006)}]{2006ApJ...641.1122B}
{Boesgaard} AM, {Novicki} MC (2006) {Beryllium in Disk and Halo Stars: Evidence
  for a Beryllium Dispersion in Old Stars}. \apj 641(2):1122--1130,
  \doi{10.1086/500501}, \eprint{astro-ph/0512317}

\bibitem[{{Boesgaard} et~al.(2011){Boesgaard}, {Rich}, {Levesque}, and
  {Bowler}}]{2011ApJ...743..140B}
{Boesgaard} AM, {Rich} JA, {Levesque} EM, {Bowler} BP (2011) {Beryllium and
  Alpha-element Abundances in a Large Sample of Metal-poor Stars}. \apj
  743(2):140, \doi{10.1088/0004-637X/743/2/140}, \eprint{1110.2823}

\bibitem[{{Bollinger} et~al.(1985){Bollinger}, {Wells}, {Wineland}, and
  {Itano}}]{1985PhRvA..31.2711B}
{Bollinger} JJ, {Wells} JS, {Wineland} DJ, {Itano} WM (1985) {Hyperfine
  structure of the 2p sup2Psub1/2 state in sup9Besup+}. \pra 31(4):2711--2714,
  \doi{10.1103/PhysRevA.31.2711}

\bibitem[{{Boyd} and {Kajino}(1989)}]{1989ApJ...336L..55B}
{Boyd} RN, {Kajino} T (1989) {Can 9Be Provide a Test of Cosmological Theories?}
  \apjl 336:L55, \doi{10.1086/185360}

\bibitem[{{Bristow} et~al.(2014){Bristow}, {Barbuy}, {Macanhan}, {Castilho},
  {Dekker}, {Delabre}, {Diaz}, {Gneiding}, {Kerber}, {Kuntschner}, {La Mura},
  {Reiss}, and {Vernet}}]{2014SPIE.9147E..09B}
{Bristow} P, {Barbuy} B, {Macanhan} VB, {Castilho} B, {Dekker} H, {Delabre} B,
  {Diaz} M, {Gneiding} C, {Kerber} F, {Kuntschner} H, {La Mura} G, {Reiss} R,
  {Vernet} J (2014) {Introducing CUBES: the Cassegrain U-band Brazil-ESO
  spectrograph}. In: {Ramsay} SK, {McLean} IS, {Takami} H (eds) Ground-based
  and Airborne Instrumentation for Astronomy V, Society of Photo-Optical
  Instrumentation Engineers (SPIE) Conference Series, vol 9147, p 914709,
  \doi{10.1117/12.2054751}

\bibitem[{{Chiba} and {Beers}(2000)}]{2000AJ....119.2843C}
{Chiba} M, {Beers} TC (2000) {Kinematics of Metal-poor Stars in the Galaxy.
  III. Formation of the Stellar Halo and Thick Disk as Revealed from a Large
  Sample of Nonkinematically Selected Stars}. \aj 119(6):2843--2865,
  \doi{10.1086/301409}, \eprint{astro-ph/0003087}

\bibitem[{{Coc} et~al.(2014){Coc}, {Uzan}, and
  {Vangioni}}]{2014JCAP...10..050C}
{Coc} A, {Uzan} JP, {Vangioni} E (2014) {Standard big bang nucleosynthesis and
  primordial CNO abundances after Planck}. \jcap 2014(10):050,
  \doi{10.1088/1475-7516/2014/10/050}, \eprint{1403.6694}

\bibitem[{{Dekker} et~al.(2000){Dekker}, {D'Odorico}, {Kaufer}, {Delabre}, and
  {Kotzlowski}}]{2000SPIE.4008..534D}
{Dekker} H, {D'Odorico} S, {Kaufer} A, {Delabre} B, {Kotzlowski} H (2000)
  {Design, construction, and performance of UVES, the echelle spectrograph for
  the UT2 Kueyen Telescope at the ESO Paranal Observatory}. In: {Iye} M,
  {Moorwood} AF (eds) Optical and IR Telescope Instrumentation and Detectors,
  Society of Photo-Optical Instrumentation Engineers (SPIE) Conference Series,
  vol 4008, pp 534--545, \doi{10.1117/12.395512}

\bibitem[{{Duncan} et~al.(1992){Duncan}, {Lambert}, and
  {Lemke}}]{1992ApJ...401..584D}
{Duncan} DK, {Lambert} DL, {Lemke} M (1992) {The Abundance of Boron in Three
  Halo Stars}. \apj 401:584, \doi{10.1086/172088}

\bibitem[{{Duncan} et~al.(1997){Duncan}, {Primas}, {Rebull}, {Boesgaard},
  {Deliyannis}, {Hobbs}, {King}, and {Ryan}}]{1997ApJ...488..338D}
{Duncan} DK, {Primas} F, {Rebull} LM, {Boesgaard} AM, {Deliyannis} CP, {Hobbs}
  LM, {King} JR, {Ryan} SG (1997) {The Evolution of Galactic Boron and the
  Production Site of the Light Elements}. \apj 488(1):338--349,
  \doi{10.1086/304683}

\bibitem[{{Emery}(2006)}]{2006sham.book..253E}
{Emery} G (2006) {Hyperfine Structure in Springer handbook of atomic,
  molecular, and optical physics}, Springer Science+Business Media, Inc., p
  253. \doi{10.1007/978-0-387-26308-3\_16}

\bibitem[{{Ernandes} et~al.(2020){Ernandes}, {Evans}, {Barbuy}, {Castilho},
  {Cescutti}, {Christlieb}, {Cristiani}, {Di Marcantonio}, {Hansen},
  {Quirrenbach}, and {Smiljanic}}]{2020SPIE11447E..60E}
{Ernandes} H, {Evans} CJ, {Barbuy} B, {Castilho} B, {Cescutti} G, {Christlieb}
  N, {Cristiani} S, {Di Marcantonio} P, {Hansen} C, {Quirrenbach} A,
  {Smiljanic} R (2020) {Stellar astrophysics in the near-UV with VLT-CUBES}.
  In: Society of Photo-Optical Instrumentation Engineers (SPIE) Conference
  Series, vol 11447, p 1144760, \doi{10.1117/12.2562497}, \eprint{2102.02205}

\bibitem[{{Ernandes} et~al.(2022){Ernandes}, {Barbuy}, {Fria{\c{c}}a}, {Hill},
  {Spite}, {Spite}, {Castilho}, and {Evans}}]{2022MNRAS.510.5362E}
{Ernandes} H, {Barbuy} B, {Fria{\c{c}}a} A, {Hill} V, {Spite} M, {Spite} F,
  {Castilho} BV, {Evans} CJ (2022) {Be, V, and Cu in the halo star CS 31082-001
  from near-UV spectroscopy}. \mnras 510(4):5362--5375,
  \doi{10.1093/mnras/stab3789}, \eprint{2202.04450}

\bibitem[{{Evans} et~al.(2016){Evans}, {Puech}, {Rodrigues}, {Barbuy}, {Cuby},
  {Dalton}, {Fitzsimons}, {Hammer}, {Jagourel}, {Kaper}, {Morris}, and
  {Morris}}]{2016SPIE.9908E..9JE}
{Evans} CJ, {Puech} M, {Rodrigues} M, {Barbuy} B, {Cuby} JG, {Dalton} G,
  {Fitzsimons} E, {Hammer} F, {Jagourel} P, {Kaper} L, {Morris} SL, {Morris} TJ
  (2016) {Science requirements and trade-offs for the MOSAIC instrument for the
  European ELT}. In: {Evans} CJ, {Simard} L, {Takami} H (eds) Ground-based and
  Airborne Instrumentation for Astronomy VI, Society of Photo-Optical
  Instrumentation Engineers (SPIE) Conference Series, vol 9908, p 99089J,
  \doi{10.1117/12.2231675}, \eprint{1608.06542}

\bibitem[{{Evans} et~al.(2018){Evans}, {Barbuy}, {Castilho}, {Smiljanic},
  {Melendez}, {Japelj}, {Cristiani}, {Snodgrass}, {Bonifacio}, {Puech}, and
  {Quirrenbach}}]{2018SPIE10702E..2EE}
{Evans} CJ, {Barbuy} B, {Castilho} B, {Smiljanic} R, {Melendez} J, {Japelj} J,
  {Cristiani} S, {Snodgrass} C, {Bonifacio} P, {Puech} M, {Quirrenbach} A
  (2018) {Revisiting the science case for near-UV spectroscopy with the VLT}.
  In: Society of Photo-Optical Instrumentation Engineers (SPIE) Conference
  Series, vol 10702, p 107022E, \doi{10.1117/12.2312022}, \eprint{1806.11173}

\bibitem[{{Fuhr} and {Wiese}(2010)}]{2010JPCRD..39a3101F}
{Fuhr} JR, {Wiese} WL (2010) {Tables of Atomic Transition Probabilities for
  Beryllium and Boron}. Journal of Physical and Chemical Reference Data
  39(1):013101--013101, \doi{10.1063/1.3286088}

\bibitem[{{Gaia Collaboration} et~al.(2016){Gaia Collaboration}, {Prusti}, {de
  Bruijne}, {Brown}, {Vallenari}, and {et al.}}]{2016A&A...595A...1G}
{Gaia Collaboration}, {Prusti} T, {de Bruijne} JHJ, {Brown} AGA, {Vallenari} A,
  {et al} (2016) {The Gaia mission}. \aap 595:A1,
  \doi{10.1051/0004-6361/201629272}, \eprint{1609.04153}

\bibitem[{{Gaia Collaboration} et~al.(2018){Gaia Collaboration}, {Babusiaux},
  {van Leeuwen}, {Barstow}, {Jordi}, {Vallenari}, and {et
  al.}}]{2018A&A...616A..10G}
{Gaia Collaboration}, {Babusiaux} C, {van Leeuwen} F, {Barstow} MA, {Jordi} C,
  {Vallenari} A, {et al} (2018) {Gaia Data Release 2. Observational
  Hertzsprung-Russell diagrams}. \aap 616:A10,
  \doi{10.1051/0004-6361/201832843}, \eprint{1804.09378}

\bibitem[{{Gilmore} et~al.(1991){Gilmore}, {Edvardsson}, and
  {Nissen}}]{1991ApJ...378...17G}
{Gilmore} G, {Edvardsson} B, {Nissen} PE (1991) {First Detection of Beryllium
  in a Very Metal Poor Star: A Test of the Standard Big Bang Model}. \apj
  378:17, \doi{10.1086/170402}

\bibitem[{{Gilmore} et~al.(1992){Gilmore}, {Gustafsson}, {Edvardsson}, and
  {Nissen}}]{1992Natur.357..379G}
{Gilmore} G, {Gustafsson} B, {Edvardsson} B, {Nissen} PE (1992) {Is beryllium
  in metal-poor stars of galactic or cosmological origin?} \nat
  357(6377):379--384, \doi{10.1038/357379a0}

\bibitem[{{Gustafsson} et~al.(2008){Gustafsson}, {Edvardsson}, {Eriksson},
  {J{\o}rgensen}, {Nordlund}, and {Plez}}]{2008A&A...486..951G}
{Gustafsson} B, {Edvardsson} B, {Eriksson} K, {J{\o}rgensen} UG, {Nordlund}
  {\r{A}}, {Plez} B (2008) {A grid of MARCS model atmospheres for late-type
  stars. I. Methods and general properties}. \aap 486(3):951--970,
  \doi{10.1051/0004-6361:200809724}, \eprint{0805.0554}

\bibitem[{{Haywood} et~al.(2018){Haywood}, {Di Matteo}, {Lehnert}, {Snaith},
  {Khoperskov}, and {G{\'o}mez}}]{2018ApJ...863..113H}
{Haywood} M, {Di Matteo} P, {Lehnert} MD, {Snaith} O, {Khoperskov} S,
  {G{\'o}mez} A (2018) {In Disguise or Out of Reach: First Clues about In Situ
  and Accreted Stars in the Stellar Halo of the Milky Way from Gaia DR2}. \apj
  863(2):113, \doi{10.3847/1538-4357/aad235}, \eprint{1805.02617}

\bibitem[{{Helmi} et~al.(2018){Helmi}, {Babusiaux}, {Koppelman}, {Massari},
  {Veljanoski}, and {Brown}}]{2018Natur.563...85H}
{Helmi} A, {Babusiaux} C, {Koppelman} HH, {Massari} D, {Veljanoski} J, {Brown}
  AGA (2018) {The merger that led to the formation of the Milky Way's inner
  stellar halo and thick disk}. \nat 563(7729):85--88,
  \doi{10.1038/s41586-018-0625-x}, \eprint{1806.06038}

\bibitem[{{Ito} et~al.(2013){Ito}, {Aoki}, {Beers}, {Tominaga}, {Honda}, and
  {Carollo}}]{2013ApJ...773...33I}
{Ito} H, {Aoki} W, {Beers} TC, {Tominaga} N, {Honda} S, {Carollo} D (2013)
  {Chemical Analysis of the Ninth Magnitude Carbon-enhanced Metal-poor Star
  BD+44{\textdegree}493}. \apj 773(1):33, \doi{10.1088/0004-637X/773/1/33},
  \eprint{1306.3614}

\bibitem[{{Korotin} and {Ku{\v{c}}inskas}(2022)}]{2022A&A...657L..11K}
{Korotin} S, {Ku{\v{c}}inskas} A (2022) {Abundance of beryllium in the Sun and
  stars: The role of non-local thermodynamic equilibrium effects}. \aap
  657:L11, \doi{10.1051/0004-6361/202142789}, \eprint{2201.00532}

\bibitem[{{Kramida}(2005)}]{2005PhyS...72..309K}
{Kramida} AE (2005) {Critical Compilation of Wavelengths and Energy Levels of
  Singly Ionized Beryllium (Be II)}. \physscr 72(4):309--319,
  \doi{10.1238/Physica.Regular.072a00309}

\bibitem[{{Krieger} et~al.(2017){Krieger}, {N{\"o}rtersh{\"a}user}, {Geppert},
  {Blaum}, {Bissell}, {Fr{\"o}mmgen}, {Hammen}, {Kreim}, {Kowalska},
  {Kr{\"a}mer}, {Neugart}, {Neyens}, {S{\'a}nchez}, {Tiedemann}, {Yordanov},
  and {Zakova}}]{2017ApPhB.123...15K}
{Krieger} A, {N{\"o}rtersh{\"a}user} W, {Geppert} C, {Blaum} K, {Bissell} ML,
  {Fr{\"o}mmgen} N, {Hammen} M, {Kreim} K, {Kowalska} M, {Kr{\"a}mer} J,
  {Neugart} R, {Neyens} G, {S{\'a}nchez} R, {Tiedemann} D, {Yordanov} DT,
  {Zakova} M (2017) {Frequency-comb referenced collinear laser spectroscopy of
  Be$^{+}$ for nuclear structure investigations and many-body QED tests}.
  Applied Physics B: Lasers and Optics 123(1):15,
  \doi{10.1007/s00340-016-6579-5}, \eprint{1609.07655}

\bibitem[{{Kusakabe} et~al.(2014){Kusakabe}, {Kim}, {Cheoun}, {Kajino}, {Kino},
  and {Mathews}}]{2014ApJS..214....5K}
{Kusakabe} M, {Kim} KS, {Cheoun} MK, {Kajino} T, {Kino} Y, {Mathews} GJ (2014)
  {Revised Big Bang Nucleosynthesis with Long-lived, Negatively Charged Massive
  Particles: Updated Recombination Rates, Primordial $^{9}$Be Nucleosynthesis,
  and Impact of New $^{6}$Li Limits}. \apjs 214(1):5,
  \doi{10.1088/0067-0049/214/1/5}, \eprint{1403.4156}

\bibitem[{{Kusakabe} et~al.(2017){Kusakabe}, {Mathews}, {Kajino}, and
  {Cheoun}}]{2017IJMPE..2641004K}
{Kusakabe} M, {Mathews} GJ, {Kajino} T, {Cheoun} MK (2017) {Review on effects
  of long-lived negatively charged massive particles on Big Bang
  Nucleosynthesis}. International Journal of Modern Physics E 26(8):1741004-64,
  \doi{10.1142/S021830131741004X}, \eprint{1706.03143}

\bibitem[{{Lingenfelter}(2019)}]{2019ApJS..245...30L}
{Lingenfelter} RE (2019) {The Origin of Cosmic Rays: How Their Composition
  Defines Their Sources and Sites and the Processes of Their Mixing, Injection,
  and Acceleration}. \apjs 245(2):30, \doi{10.3847/1538-4365/ab4b58},
  \eprint{1903.06330}

\bibitem[{{Mathar}(2011)}]{2011arXiv1102.5125M}
{Mathar} RJ (2011) {Corrigendum to ``Universal factorization of 3n-j (j>2)
  symbols...'' [J. Phys. A: Math. Gen. 37 (2004) 3259]}. arXiv e-prints
  arXiv:1102.5125, \eprint{1102.5125}

\bibitem[{{Molaro} and {Beckman}(1984)}]{1984A&A...139..394M}
{Molaro} P, {Beckman} J (1984) {An upper limit to the abundance of 9Be in the
  population II star HD 76932 from a high resolution spectrum with the IUE.}
  \aap 139:394--400

\bibitem[{{Molaro} et~al.(1984){Molaro}, {Beckman}, and
  {Castelli}}]{1984ESASP.218..197M}
{Molaro} P, {Beckman} JE, {Castelli} F (1984) {An upper limit to the beryllium
  abundance in the population II star HD 140283 from a high resolution IUE
  spectrum.} In: {Rolfe} E (ed) Fourth European IUE Conference, ESA Special
  Publication, vol 218, pp 197--201

\bibitem[{{Molaro} et~al.(2020){Molaro}, {Cescutti}, and
  {Fu}}]{2020MNRAS.496.2902M}
{Molaro} P, {Cescutti} G, {Fu} X (2020) {Lithium and beryllium in the
  Gaia-Enceladus galaxy}. \mnras 496(3):2902--2909,
  \doi{10.1093/mnras/staa1653}, \eprint{2006.00787}

\bibitem[{{N{\"o}rtersh{\"a}user} et~al.(2015){N{\"o}rtersh{\"a}user},
  {Geppert}, {Krieger}, {Pachucki}, {Puchalski}, {Blaum}, {Bissell},
  {Fr{\"o}mmgen}, {Hammen}, {Kowalska}, {Kr{\"a}mer}, {Kreim}, {Neugart},
  {Neyens}, {S{\'a}nchez}, and {Yordanov}}]{2015PhRvL.115c3002N}
{N{\"o}rtersh{\"a}user} W, {Geppert} C, {Krieger} A, {Pachucki} K, {Puchalski}
  M, {Blaum} K, {Bissell} ML, {Fr{\"o}mmgen} N, {Hammen} M, {Kowalska} M,
  {Kr{\"a}mer} J, {Kreim} K, {Neugart} R, {Neyens} G, {S{\'a}nchez} R,
  {Yordanov} DT (2015) {Precision Test of Many-Body QED in the Be$^{+}$ 2 p
  Fine Structure Doublet Using Short-Lived Isotopes}. \prl 115(3):033002,
  \doi{10.1103/PhysRevLett.115.033002}, \eprint{1507.03830}

\bibitem[{{Orito} et~al.(1997){Orito}, {Kajino}, {Boyd}, and
  {Mathews}}]{1997ApJ...488..515O}
{Orito} M, {Kajino} T, {Boyd} RN, {Mathews} GJ (1997) {Geometrical Effects of
  Baryon Density Inhomogeneities on Primordial Nucleosynthesis}. \apj
  488(2):515--523, \doi{10.1086/304716}, \eprint{astro-ph/9609130}

\bibitem[{{Pasquini}(2014)}]{2014Ap&SS.354..121P}
{Pasquini} L (2014) {UV opportunities at ESO}. \apss 354(1):121--124,
  \doi{10.1007/s10509-014-2049-x}

\bibitem[{{Pasquini} et~al.(2004){Pasquini}, {Bonifacio}, {Randich}, {Galli},
  and {Gratton}}]{2004A&A...426..651P}
{Pasquini} L, {Bonifacio} P, {Randich} S, {Galli} D, {Gratton} RG (2004)
  {Beryllium in turnoff stars of NGC 6397: Early Galaxy spallation,
  cosmochronology and cluster formation.} \aap 426:651--657,
  \doi{10.1051/0004-6361:20041254}, \eprint{astro-ph/0407524}

\bibitem[{{Pasquini} et~al.(2005){Pasquini}, {Galli}, {Gratton}, {Bonifacio},
  {Randich}, and {Valle}}]{2005A&A...436L..57P}
{Pasquini} L, {Galli} D, {Gratton} RG, {Bonifacio} P, {Randich} S, {Valle} G
  (2005) {Early star formation in the Galaxy from beryllium and oxygen
  abundances}. \aap 436(3):L57--L60, \doi{10.1051/0004-6361:200500124},
  \eprint{astro-ph/0505396}

\bibitem[{{Pasquini} et~al.(2007){Pasquini}, {Bonifacio}, {Randich}, {Galli},
  {Gratton}, and {Wolff}}]{2007A&A...464..601P}
{Pasquini} L, {Bonifacio} P, {Randich} S, {Galli} D, {Gratton} RG, {Wolff} B
  (2007) {Beryllium abundance in turn-off stars of NGC 6752}. \aap
  464(2):601--607, \doi{10.1051/0004-6361:20066260}, \eprint{astro-ph/0612710}

\bibitem[{{Placco} et~al.(2014){Placco}, {Beers}, {Roederer}, {Cowan},
  {Frebel}, {Filler}, {Ivans}, {Lawler}, {Schatz}, {Sneden}, {Sobeck}, {Aoki},
  and {Smith}}]{2014ApJ...790...34P}
{Placco} VM, {Beers} TC, {Roederer} IU, {Cowan} JJ, {Frebel} A, {Filler} D,
  {Ivans} II, {Lawler} JE, {Schatz} H, {Sneden} C, {Sobeck} JS, {Aoki} W,
  {Smith} VV (2014) {Hubble Space Telescope Near-ultraviolet Spectroscopy of
  the Bright CEMP-no Star BD+44{\textdegree}493}. \apj 790(1):34,
  \doi{10.1088/0004-637X/790/1/34}, \eprint{1406.0538}

\bibitem[{{Plez}(2012)}]{2012ascl.soft05004P}
{Plez} B (2012) {Turbospectrum: Code for spectral synthesis}. \eprint{1205.004}

\bibitem[{{Pospelov} et~al.(2008){Pospelov}, {Pradler}, and
  {Steffen}}]{2008JCAP...11..020P}
{Pospelov} M, {Pradler} J, {Steffen} FD (2008) {Constraints on supersymmetric
  models from catalytic primordial nucleosynthesis of beryllium}. \jcap
  2008(11):020, \doi{10.1088/1475-7516/2008/11/020}, \eprint{0807.4287}

\bibitem[{{Prantzos}(2012)}]{2012A&A...542A..67P}
{Prantzos} N (2012) {Production and evolution of Li, Be, and B isotopes in the
  Galaxy}. \aap 542:A67, \doi{10.1051/0004-6361/201219043}, \eprint{1203.5662}

\bibitem[{{Primas} et~al.(2000{\natexlab{a}}){Primas}, {Asplund}, {Nissen}, and
  {Hill}}]{2000A&A...364L..42P}
{Primas} F, {Asplund} M, {Nissen} PE, {Hill} V (2000{\natexlab{a}}) {The
  beryllium abundance in the very metal-poor halo star G 64-12 from VLT/UVES
  observations}. \aap 364:L42--L46, \eprint{astro-ph/0009482}

\bibitem[{{Primas} et~al.(2000{\natexlab{b}}){Primas}, {Molaro}, {Bonifacio},
  and {Hill}}]{2000A&A...362..666P}
{Primas} F, {Molaro} P, {Bonifacio} P, {Hill} V (2000{\natexlab{b}}) {First
  UVES observations of beryllium in very metal-poor stars}. \aap 362:666--672,
  \eprint{astro-ph/0008402}

\bibitem[{{Puchalski} and {Pachucki}(2009)}]{2009PhRvA..79c2510P}
{Puchalski} M, {Pachucki} K (2009) {Fine and hyperfine splitting of the 2P
  state in Li and Be$^{+}$}. \pra 79(3):032510,
  \doi{10.1103/PhysRevA.79.032510}, \eprint{0901.2633}

\bibitem[{{Puchalski} and {Pachucki}(2014)}]{2014PhRvA..89c2510P}
{Puchalski} M, {Pachucki} K (2014) {Ground-state hyperfine splitting in the
  Be$^{+}$ ion}. \pra 89(3):032510, \doi{10.1103/PhysRevA.89.032510},
  \eprint{1402.4573}

\bibitem[{{Puchalski} and {Pachucki}(2015)}]{2015PhRvA..92a2513P}
{Puchalski} M, {Pachucki} K (2015) {Quantum electrodynamics m
  {\ensuremath{\alpha}}$^{6}$ and m
  {\ensuremath{\alpha}}$^{7}$ln{\ensuremath{\alpha}} corrections to the fine
  splitting in Li and Be$^{+}$}. \pra 92(1):012513,
  \doi{10.1103/PhysRevA.92.012513}, \eprint{1506.02462}

\bibitem[{{Rebolo} et~al.(1988){Rebolo}, {Molaro}, {Abia}, and
  {Beckman}}]{1988A&A...193..193R}
{Rebolo} R, {Molaro} P, {Abia} C, {Beckman} JE (1988) {Abundances of 9Be in a
  sample of highly metal-deficient dwarfs : implications for early galactic
  nucleosynthesis and primordial lithium.} \aap 193:193--201

\bibitem[{{Rich} and {Boesgaard}(2009)}]{2009ApJ...701.1519R}
{Rich} JA, {Boesgaard} AM (2009) {Beryllium, Oxygen, and Iron Abundances in
  Extremely Metal-Deficient Stars}. \apj 701(2):1519--1533,
  \doi{10.1088/0004-637X/701/2/1519}, \eprint{0906.3296}

\bibitem[{{Ryan} et~al.(1992){Ryan}, {Norris}, {Bessell}, and
  {Deliyannis}}]{1992ApJ...388..184R}
{Ryan} SG, {Norris} JE, {Bessell} MS, {Deliyannis} C (1992) {Evolution of
  Beryllium Abundances in the Galactic Halo}. \apj 388:184,
  \doi{10.1086/171141}

\bibitem[{{Safronova} and {Safronova}(2013)}]{2013PhRvA..87c2502S}
{Safronova} UI, {Safronova} MS (2013) {Relativistic many-body calculation of
  energies, lifetimes, polarizabilities, and hyperpolarizabilities in Li-like
  Be$^{+}$}. \pra 87(3):032502, \doi{10.1103/PhysRevA.87.032502}

\bibitem[{{Shukla} et~al.(2020){Shukla}, {Arora}, {Sharma}, and
  {Srivastava}}]{2020PhRvA.102b2817S}
{Shukla} N, {Arora} B, {Sharma} L, {Srivastava} R (2020) {Two-dipole and
  three-dipole dispersion coefficients for interaction of alkaline-earth-metal
  atoms with alkaline-earth-metal atoms and alkaline-earth-metal ions}. \pra
  102(2):022817, \doi{10.1103/PhysRevA.102.022817}, \eprint{2008.04341}

\bibitem[{{Smiljanic}(2014)}]{2014Ap&SS.354...55S}
{Smiljanic} R (2014) {Stellar abundances of beryllium and CUBES}. \apss
  354(1):55--64, \doi{10.1007/s10509-014-1916-9}, \eprint{1403.6276}

\bibitem[{{Smiljanic} et~al.(2009){Smiljanic}, {Pasquini}, {Bonifacio},
  {Galli}, {Gratton}, {Randich}, and {Wolff}}]{2009A&A...499..103S}
{Smiljanic} R, {Pasquini} L, {Bonifacio} P, {Galli} D, {Gratton} RG, {Randich}
  S, {Wolff} B (2009) {Beryllium abundances and star formation in the halo and
  in the thick disk}. \aap 499(1):103--119, \doi{10.1051/0004-6361/200810592},
  \eprint{0902.0483}

\bibitem[{{Smiljanic} et~al.(2021){Smiljanic}, {Zych}, and
  {Pasquini}}]{2021A&A...646A..70S}
{Smiljanic} R, {Zych} MG, {Pasquini} L (2021) {Inhomogeneity in the early
  Galactic chemical enrichment exposed by beryllium abundances in extremely
  metal-poor stars}. \aap 646:A70, \doi{10.1051/0004-6361/202039101},
  \eprint{2012.07438}

\bibitem[{{Soubiran} et~al.(2016){Soubiran}, {Le Campion}, {Brouillet}, and
  {Chemin}}]{2016A&A...591A.118S}
{Soubiran} C, {Le Campion} JF, {Brouillet} N, {Chemin} L (2016) {The PASTEL
  catalogue: 2016 version}. \aap 591:A118, \doi{10.1051/0004-6361/201628497},
  \eprint{1605.07384}

\bibitem[{{Spite} et~al.(2019){Spite}, {Bonifacio}, {Spite}, {Caffau},
  {Sbordone}, and {Gallagher}}]{2019A&A...624A..44S}
{Spite} M, {Bonifacio} P, {Spite} F, {Caffau} E, {Sbordone} L, {Gallagher} AJ
  (2019) {Be and O in the ultra metal-poor dwarf 2MASS J18082002-5104378: the
  Be-O correlation}. \aap 624:A44, \doi{10.1051/0004-6361/201834741},
  \eprint{1902.11048}

\bibitem[{{Suda} et~al.(2008){Suda}, {Katsuta}, {Yamada}, {Suwa}, {Ishizuka},
  {Komiya}, {Sorai}, {Aikawa}, and {Fujimoto}}]{2008PASJ...60.1159S}
{Suda} T, {Katsuta} Y, {Yamada} S, {Suwa} T, {Ishizuka} C, {Komiya} Y, {Sorai}
  K, {Aikawa} M, {Fujimoto} MY (2008) {Stellar Abundances for the Galactic
  Archeology (SAGA) Database --- Compilation of the Characteristics of Known
  Extremely Metal-Poor Stars}. \pasj 60:1159, \doi{10.1093/pasj/60.5.1159},
  \eprint{0806.3697}

\bibitem[{{Suzuki} and {Yoshii}(2001)}]{2001ApJ...549..303S}
{Suzuki} TK, {Yoshii} Y (2001) {A New Model for the Evolution of Light Elements
  in an Inhomogeneous Galactic Halo}. \apj 549(1):303--319,
  \doi{10.1086/319049}, \eprint{astro-ph/0010108}

\bibitem[{{Suzuki} et~al.(1999){Suzuki}, {Yoshii}, and
  {Kajino}}]{1999ApJ...522L.125S}
{Suzuki} TK, {Yoshii} Y, {Kajino} T (1999) {Evolution of Beryllium and Boron in
  the Inhomogeneous Early Galaxy}. \apjl 522(2):L125--L128,
  \doi{10.1086/312233}, \eprint{astro-ph/9907182}

\bibitem[{{Tan} and {Zhao}(2011)}]{2011ApJ...738L..33T}
{Tan} K, {Zhao} G (2011) {A Possible Signature of Non-uniform
  Be-{\ensuremath{\alpha}} Relationships for the Galaxy}. \apjl 738(2):L33,
  \doi{10.1088/2041-8205/738/2/L33}, \eprint{1108.2074}

\bibitem[{{Tan} et~al.(2009){Tan}, {Shi}, and {Zhao}}]{2009MNRAS.392..205T}
{Tan} KF, {Shi} JR, {Zhao} G (2009) {Beryllium abundances in metal-poor stars}.
  \mnras 392(1):205--215, \doi{10.1111/j.1365-2966.2008.14027.x},
  \eprint{0810.2600}

\bibitem[{{Tang} et~al.(2009){Tang}, {Zhang}, {Yan}, {Shi}, {Babb}, and
  {Mitroy}}]{2009PhRvA..80d2511T}
{Tang} LY, {Zhang} JY, {Yan} ZC, {Shi} TY, {Babb} JF, {Mitroy} J (2009)
  {Calculations of polarizabilities and hyperpolarizabilities for the Be$^{+}$
  ion}. \pra 80(4):042511, \doi{10.1103/PhysRevA.80.042511}, \eprint{0908.4060}

\bibitem[{{Tatischeff} and {Gabici}(2018)}]{2018ARNPS..68..377T}
{Tatischeff} V, {Gabici} S (2018) {Particle Acceleration by Supernova Shocks
  and Spallogenic Nucleosynthesis of Light Elements}. Annual Review of Nuclear
  and Particle Science 68(1):377--404,
  \doi{10.1146/annurev-nucl-101917-021151}, \eprint{1803.01794}

\bibitem[{{Valle} et~al.(2002){Valle}, {Ferrini}, {Galli}, and
  {Shore}}]{2002ApJ...566..252V}
{Valle} G, {Ferrini} F, {Galli} D, {Shore} SN (2002) {Evolution of Li, Be, and
  B in the Galaxy}. \apj 566(1):252--260, \doi{10.1086/338036},
  \eprint{astro-ph/0110327}

\bibitem[{Woodgate(1970)}]{woodgate1970elementary}
Woodgate GK (1970) Elementary Atomic Structure. Oxford University Press

\bibitem[{{Yan} et~al.(1998){Yan}, {Tambasco}, and
  {Drake}}]{1998PhRvA..57.1652Y}
{Yan} ZC, {Tambasco} M, {Drake} GWF (1998) {Energies and oscillator strengths
  for lithiumlike ions}. \pra 57(3):1652--1661, \doi{10.1103/PhysRevA.57.1652}

\bibitem[{{Yerokhin}(2008)}]{2008PhRvA..78a2513Y}
{Yerokhin} VA (2008) {Hyperfine structure of Li and Be$^{+}$}. \pra
  78(1):012513, \doi{10.1103/PhysRevA.78.012513}, \eprint{0805.0677}

\end{thebibliography}

%
%

\end{document}